\documentclass{article}
\usepackage{amsmath,amsfonts,amssymb,graphicx,epstopdf,cancel,bm}
\pdfoutput=1
\newcommand{\ud}{\mathrm{d}}
\newcommand{\ue}{\mathrm{e}}
\newcommand{\ui}{\mathrm{i}}
\newcommand{\uu}{\mathcal{U}}
\title{Optimal transient growth in an incompressible flow past a backward-slanted step}
\author{Marco Martins Afonso$^{1,2,*}$, Philippe Meliga$^2$ \& Eric Serre$^2$\\[0.2cm]
 {\footnotesize$^1$ Centro de Matem\'atica (Faculdade de Ci\^encias) da Universidade do Porto,}\\[-0.1cm]{\footnotesize Rua do Campo Alegre 687, 4169-007 Porto, Portugal}\\[0.2cm]
 {\footnotesize$^2$ Aix Marseille Universit\'e, CNRS, Centrale Marseille, M2P2 UMR 7340,}\\[-0.1cm]{\footnotesize 13451 Marseille, France}}
\date{* marcomartinsafonso@hotmail.it}

\begin{document}

 \maketitle

 \begin{abstract}
  With the aim of providing a first step in the quest for a reduction of the aerodynamic drag on the rear-end of a car,
  we study the phenomena of separation and reattachment of an incompressible flow focusing on a specific aerodynamic geometry,
  namely a backward-slanted step at 25$^{\circ}$ of inclination. The ensuing recirculation bubble provides the basis for an analytical and numerical
  investigation of streamwise-streak generation, lift-up effect, and turbulent-wake and Kelvin--Helmholtz instabilities.
  A linear stability analysis is performed, and an optimal control problem with a steady volumic forcing is tackled by means of variational formulation,
  adjoint method, penalization scheme and orthogonalization algorithm. Dealing with the transient growth of spanwise-periodic perturbations and inspired
  by the need of physically-realizable disturbances, we finally provide a procedure attaining a kinetic-energy maximal gain of the order of $10^6$
  with respect to the power introduced by the external forcing.
 \end{abstract}


 \section{Introduction}
 
 The research field of hydrodynamic stability has the objective of elucidating how structures of some specific temporal frequency and spatial scale
 are selected and emerge owing to the amplification of small-magnitude perturbations. The comprehension of these effects is of huge relevance,
 since many flows of practical interest are dominated by genuine instability mechanisms that can be either enhanced or alleviated to improve performances.
 Typical expected benefits consist in the reduction of the operational cost of vehicles by decreasing the skin friction or the aerodynamic drag,
 or the extension of the operating conditions of turbomachinery by increasing the surface heat flux.
 The investigation is based on structural sensitivity \cite{GL07,MSJ08}, a theoretical concept stemming from the framework of stability analysis
 in laminar flows. This allows one to identify beforehand which regions of a given flow are most sensitive to a prescribed actuation,
 without the needs of calculating the actual controlled flow and of resorting systematically to a trial-and-error procedure,
 which would represent an insurmountable bottleneck. Here we apply this concept to determine where and how to control
 efficiently the turbulent-flow separation occurring at the rear-end of a ground vehicle.
 Such an approach can thus be used to obtain valuable information about the most sensitive regions for open-loop control based on the underlying physics. 

 Separated flows often arise in industrial applications, as results of an adverse pressure gradient stemming from either operating
 conditions or geometrical constraints (airfoil at high angles of attack, rear-end of a blunt body).
 They are usually associated with a loss of performance.
 For a ground vehicle, the flow separation taking place at its rear-end contributes to a huge increase in the drag force,
 and thus also in the fuel consumption and pollutant emissions.
 For instance, a drag increase by 10\% is expected to augment the fuel consumption by 5\% at highway speeds.
 Moreover, flow separation causes low-frequency instabilities, which can trigger the excitation of aeroacoustic
 noise (sunroof cavities, side mirrors). The implementation of efficient control strategies, aimed at preventing separation itself or ---
 when this is prohibitively costly or inevitable --- at alleviating its detrimental consequences, is therefore a great environmental and economical issue.
 The dynamics is here triggered by complex interactions between small-scale structures inside the shear layer, huge flow separations,
 and trailing vortices expanding far in the wake.
 
 Many complex phenomena are attacked by means of linear perturbation dynamics,
 aiming at describing the fate of infinitesimal disturbances superimposed on a
 steady basic flow, and providing a rigorous mathematical foundation to investigate the control of fluid systems.
 Various perspectives have emerged, depending on whether the disturbance growth is characterized at large or short times: on the one hand,
 the archetype of disturbance energy amplified over asymptotically large times is the occurrence of vortex-shedding in wake flows,
 a behaviour called a \emph{modal} instability \cite{GL07}; on the other hand, the transient amplification over finite times is typically observed in
 channel flows, and is referred to as a \emph{non-modal} instability \cite{S07,AEGH08,SMMB10}.

 For boundary-layer-like flows exhibiting a marginal separation, as occurring at the rear-end of a vehicle with small slant angle,
 a non-modal theoretical analysis can identify flow regions where the transient amplification of streamwise streaks (by the lift-up effect) is most
 sensitive to steady spanwise periodic disturbances \cite{CPD09}.
 In the experiments, such disturbances can result from either steady jets or roughness elements
 positioned upstream of the separation location, reproducing a parietal or a volumic forcing respectively.
 Indeed, it is very-well-known that the global dynamics of complex flows can be modified by imposing local disturbances.
 Typical examples are the use of surface rugosities to delay the transition to turbulence in boundary layers,
 or the injection of fluid into the wake of a bluff body to alleviate unsteadiness \cite{PCD10}.
 We remind that the \emph{lift-up} effect is related to the vertical mixing of large-speed fluid from higher layers to lower ones,
 and vice versa for small-speed fluid. A streamwise vorticity perturbation arises, and evolves into a set of streamwise stripes
 characterized by relevant variations of the streamwise velocity, possibly with a periodic structure in the spanwise direction: the \emph{streaks}.

 In practice, one can use the adjoint method to calculate the gradient of some objective function (energy gain over a specified time horizon,
 growth rate of unstable disturbances) with respect to each actuation parameter, thus making it possible to cover large parameter
 spaces with a limited number of computations. This capability is useful as an aid to guide the design of efficient, tractable control strategies.
 In the past, it has mainly been applied to related problems of vortex shedding in compressible or incompressible laminar wakes,
 and the agreement between the experimental results and the theoretical predictions is excellent near the threshold of instability.
 Its application to flow separation for ground vehicles of practical importance constitutes a major issue,
 since substantial developments are necessary in order to encompass the complexity of
 turbulent-flow regimes, where large intervals of temporal and spatial scales strongly interact.

 The paper is organized as follows. In section~\ref{val} we describe
 our numerical approach and its validation. In section~\ref{geo} we
 specify the geometry under consideration and the main equations into play.
 In section~\ref{bas} we introduce the base flow we have adopted. In section~\ref{lsa}
 we perform the linear-stability analysis and we focus on the direct and adjoint
 perturbations. In section~\ref{con} we analyze the control mechanisms and the
 associated kinetic-energy gain. Conclusions and perspectives follow in section~\ref{per}.
 The appendix~\ref{app} is devoted to showing some further details about boundary conditions and adjoint equations.

 \section{Description and validation of numerical tools} \label{val}

 We have made use of the FreeFEM++ software \cite{H12} to build a Finite-Element Method code.
 Such a tool solves the continuity and Navier--Stokes equations in their variational formulation,
 with prescribed boundary conditions (Dirichlet, Neumann or mixed).
 We have implemented a P1 scheme for the pressure field, and a P1b scheme for the velocity field
 (for which we have also tested a P2 scheme, without any appreciable change in the results) \cite{MG11}.
 
 We have performed two main validation tests. In both cases, the quantitative validation makes use of a software which,
 by scanning printed figures from scientific articles, gives the numerical values of plotted points or lines with great precision.\\
 \begin{figure}[t]
  \centering
  \includegraphics[scale=0.15]{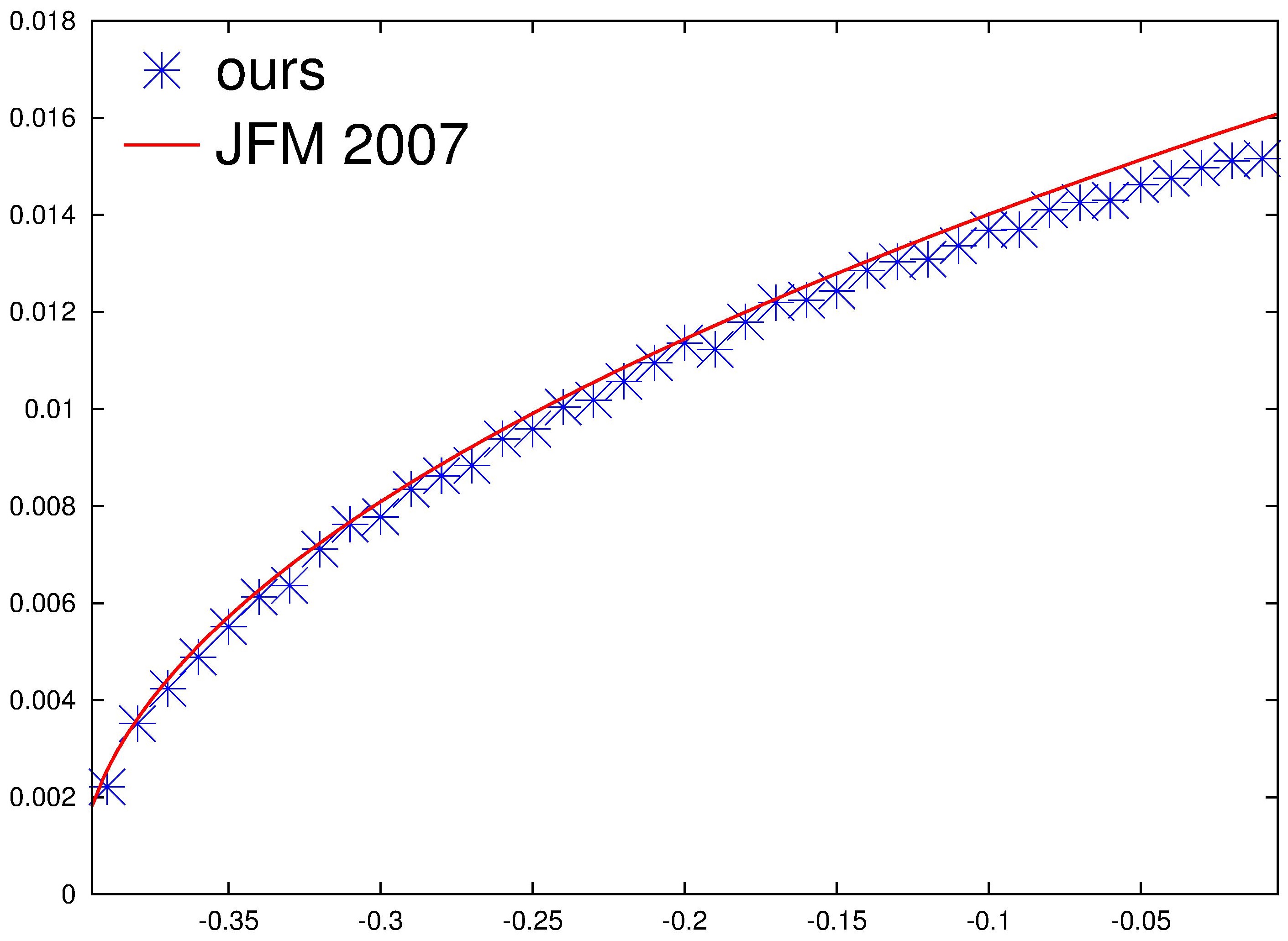}
  \caption{Validation against figure 9b of reference \cite{SL07}. Plotted in ordinate is the generalized displacement thickness (\ref{gdt})
   versus the streamwise coordinate (in abscissa).}
  \label{acc}
 \end{figure}%
 First, we have considered the 2D open cavity from \cite{SL07}. We have implemented the same exact geometry and mesh from figure 7 thereof,
 consisting of a long flat floor interrupted by a unit square excavated below it.
 We have performed a qualitative validation against their figures 8c\&d and 10, for the direct \& adjoint perturbation and the eigenvalues.
 More importantly, we have focused on their figure 9b reporting the generalized displacement thickness (to be defined more precisely in
 (\ref{gdt})), and we have found a good quantitative agreement as shown in our figure~\ref{acc}.
 (Notice that the prefactor $2.71$ reported in the caption of their figure 9b in \cite{SL07} is incorrect,
 as the correct coefficient from the theory and in the plot is actually $1.72$ \cite{EG05}.)\\
 \begin{figure}[t]
  \centering
  \includegraphics[scale=0.15]{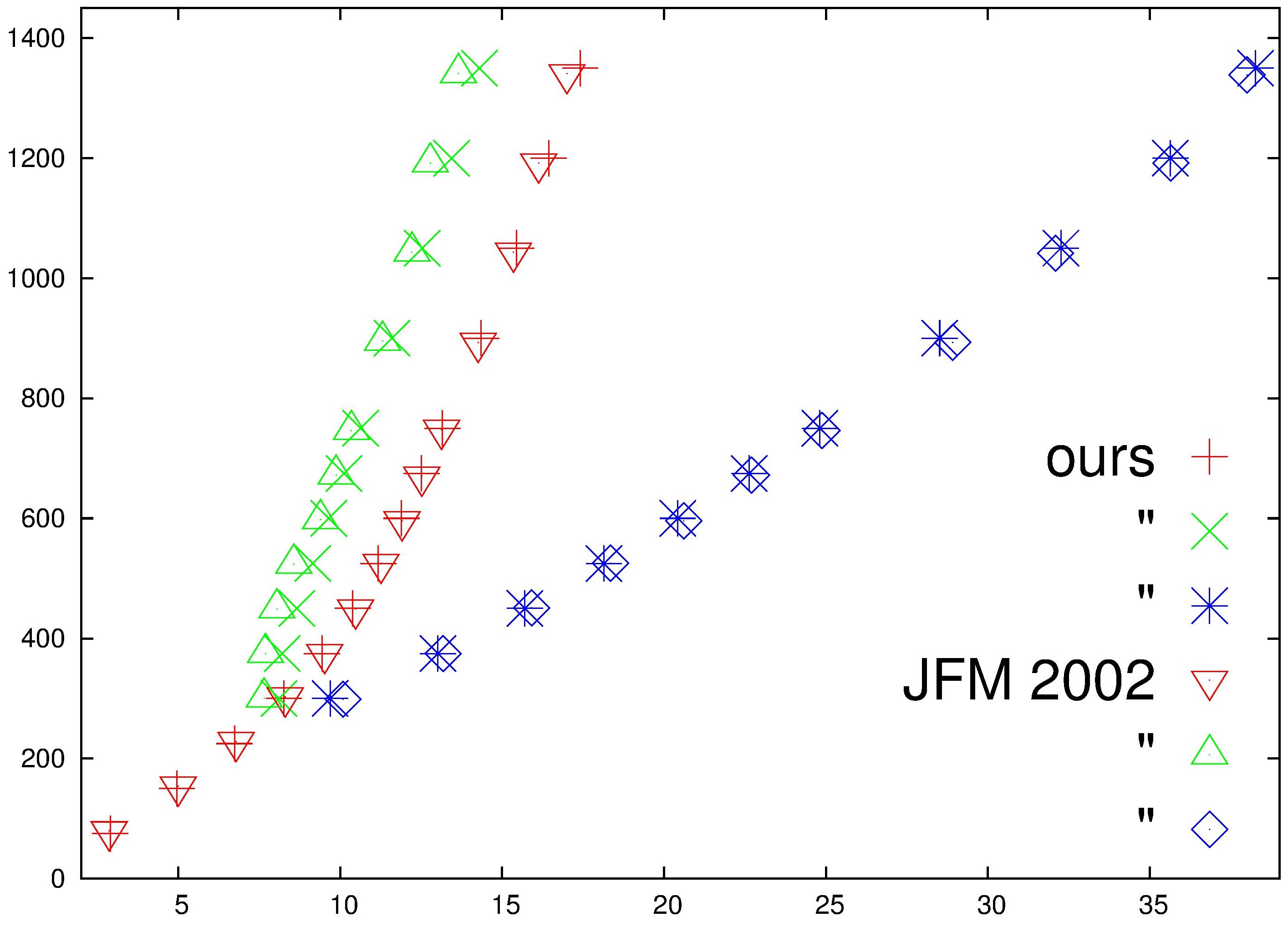}
  \caption{Validation against figure 5 of reference \cite{BGH01}. As functions of the Reynolds number (in ordinates), plotted are the abscissae of the
   reattachment point on the lower wall $\forall\,\mathrm{Re}$ (red plus signs and inverted triangles), and --- only for $\mathrm{Re}\ge300$ --- of the
   separation (green times signs and standard triangles) and reattachment (blue stars and diamonds) points on the upper wall.}
  \label{asm}
 \end{figure}%
 Second, we have considered the 3D backward-facing step (inclined at $90^{\circ}$) from \cite{BGH01}.
 We have implemented the same exact geometry from their figure 1, and (because of the different numerical scheme) an approximate mesh from their figure 2.
 We have performed a qualitative validation against their figures 3-4-7 for the base flow, the skin friction, and the eigenvalues.
 More importantly, we have focused on their figure 5 reporting the separation/reattachment points,
 and we have found a good quantitative agreement as shown in our figure~\ref{asm}.

 \section{Geometry and equations} \label{geo}
 
\begin{figure}
  \centering
  \includegraphics[scale=0.9]{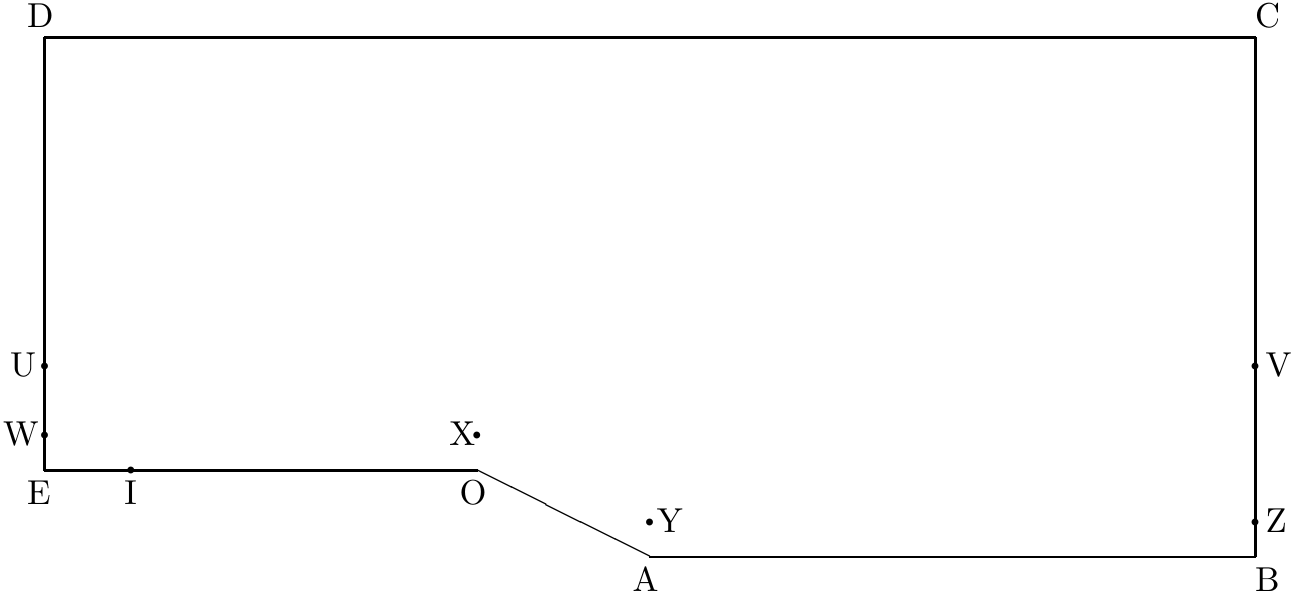}
  \caption{Our reference geometry, with the following point coordinates (notice that the figure is not to scale). Physical points: O=$(0,0)$,
   A=$(2.1445,-1)$, B=$(100,-1)$, C=$(100,30)$, D=$(-25,30)$, E=$(-25,0)$. Only for boundary conditions: I=$(-20,0)$. Only for meshing:
   U=$(-25,0.5)$, V=$(100,0.5)$, W=$(-25,0.1)$, X=$(0,0.1)$, Y=$(2.1445,-0.9)$, Z=$(100,-0.9)$. The $x$ axis points to the right and the $y$ axis
   to the top, with invariance with respect to the $z$ axis. The length of the sloping portion is $\overline{\text{OA}}=2.3662$.}
   \label{fig}
 \end{figure}

 We have focused on a geometry issued from a standard ERCOFTAC benchmark,
 namely a backward-slanted step with a slope of $25^{\circ}$ with respect to the horizontal surface.
 This configuration, plotted in figure~\ref{fig}, represents a simplification of the rear end of a car
 and of a portion of Ahmed's body \cite{AR84}.
 The $x$ and $y$ axes correspond to the streamwise and wall-normal components, respectively,
 with the origin placed at the leftmost/uppermost point of the sloping zone.
 This picture is assumed as invariant in the spanwise $z$ direction,
 which implies that the base flow is assumed as two-dimensional,
 while the perturbations can present a three-dimensional character.
 (Different geometries investigated through this scheme can be found e.g.\ in \cite{SJ00,GME07,MC11}.)\\
 The linear density of meshing points for the automatic triangulation process has been assumed as 4 on segments UD, DC, CV; 14 on segments WU, UV, VZ;
 and 24 on segments EW, WX, XY, YZ, ZB, BA, AO, OI, IE --- a finer grid is obviously required close to the lower physical boundary.
 The resulting number of triangular elements employed in the numerical simulations is about $5\times10^5$.\\
 About the nondimensionalization, we have assumed as reference units the vertical projection of the step
 and the uniform inlet speed. As the two quantities have unitary values for us,
 the nondimensionalized kinetic viscosity $\nu$ equals the inverse of the Reynolds number based on the step height.\\
 We have imposed standard inlet and outlet conditions on the left and right boundaries, respectively,
 and a free-slip condition on the upper boundary. For what concerns the lower boundary (a physical wall),
 we have imposed the no-slip condition, except for the very first portion EI on which a free-slip condition
 has been used, in order to allow for the evolution of a boundary-layer profile \cite{BGP11}. (See also appendix~\ref{app}.)\\
 Tests have also been made in which we have varied the streamwise length of the domain (both upstream and downstream), its normal height,
 and the length of the segment EI for the imposition of the boundary condition. The chosen reference geometry falls
 in a range where convergence has already taken place. The case in which the upper boundary is a physical wall has also been briefly
 investigated, both in the case of the standard segment DC, and in a modified domain where the latter has a curved S-shaped profile to simulate
 a streamline and study the influence of confinement \cite{MSCJ08,MLCSJ09}. The effect of the resolution has been tested as well,
 by implementing a discretization of up to almost $7\times10^5$ triangles, without appreciable changes.

 The full incompressible flow $\displaystyle\binom{\bm{u}}{p}(x,y,z,t)$, comprising both the velocity and the pressure fields,
 satisfies the Navier--Stokes and continuity equations,
 \begin{equation*}
  \left\{\begin{array}{rcl}\partial_t\bm{u}+\bm{u}\cdot\bm{\nabla}\bm{u}\!\!&\!\!=\!\!&\!\!-\bm{\nabla}p+\nu\nabla^2\bm{u}\;,\\
  \bm{\nabla}\cdot\bm{u}\!\!&\!\!=\!\!&\!\!0\;.\end{array}\right.
 \end{equation*}
 (For an interesting discussion on the role of compressibility --- not considered here --- see e.g.\ \cite{MSC10a,MSC10b}.)
 In what follows, we decompose it into a 2D steady solution plus a 3D small perturbation:
 \begin{equation} \label{dec}
  \binom{\bm{u}}{p}=\binom{\bm{U}}{P}+\binom{\bm{u}'}{p'}\;,\ \textrm{with}\ \bm{U}=\left(\begin{array}{c}U\\V\\0\end{array}\right)(x,y)\;,\ \textrm{and}\ \bm{u}'=\left(\begin{array}{c}u'\\v'\\w'\end{array}\right)(x,y,z,t)\;,
 \end{equation}
 for $|\bm{u}'|\ll|\bm{U}|$ and $|p'|\ll|P|$.

 \section{Base flow} \label{bas}
 
 We have assumed as our base flow, $\displaystyle\binom{\bm{U}}{P}$, a steady solution of the Navier--Stokes and continuity equations,
 \begin{equation*}
  \left\{\begin{array}{rcl}\bm{U}\cdot\bm{\nabla}\bm{U}\!\!&\!\!=\!\!&\!\!-\bm{\nabla}P+\nu\nabla^2\bm{U}\;,\\
  \bm{\nabla}\cdot\bm{U}\!\!&\!\!=\!\!&\!\!0\;,\end{array}\right.
 \end{equation*}
 satisfying the same boundary conditions as the full flow. We have obtained this flow numerically by means of Newton's iterative method \cite{F85}.
 (The relevance of small modifications in the base flow was studied e.g.\ in \cite{BCL03,BSPM11}.)
 
 Notice that, because of mass conservation, this type of base flow presents a speed overshoot,
 i.e.\ for some range of $y$ the horizontal velocity exceeds unity.
 The vertical profile is not monotonic as in a standard boundary layer, so that it is not appropriate
 to define a typical width as the height at which the velocity reaches a definite percentage of the far-field value.
 It is therefore more convenient to quantify the boundary layer by means of the so-called ``generalized displacement thickness'' \cite{SL07}:
 \begin{equation} \label{gdt}
  \delta_1(x)\equiv\frac{\int\ud y\,y\,\omega(x,y)}{\int\ud y\,\omega(x,y)}\;,
 \end{equation}
 where $\omega\equiv\partial_xV-\partial_yU$ represents the vorticity of the base flow
 (a scalar quantity, i.e.\ the $z$ component --- the only non-zero --- of the vector given by the curl of the base velocity).
 When this profile reaches the step, we find $\delta_1(x=0)\in[0.08,0.18]$, depending on the Reynolds number.
 
\begin{figure}[t]
  \centering
  \includegraphics[scale=0.15]{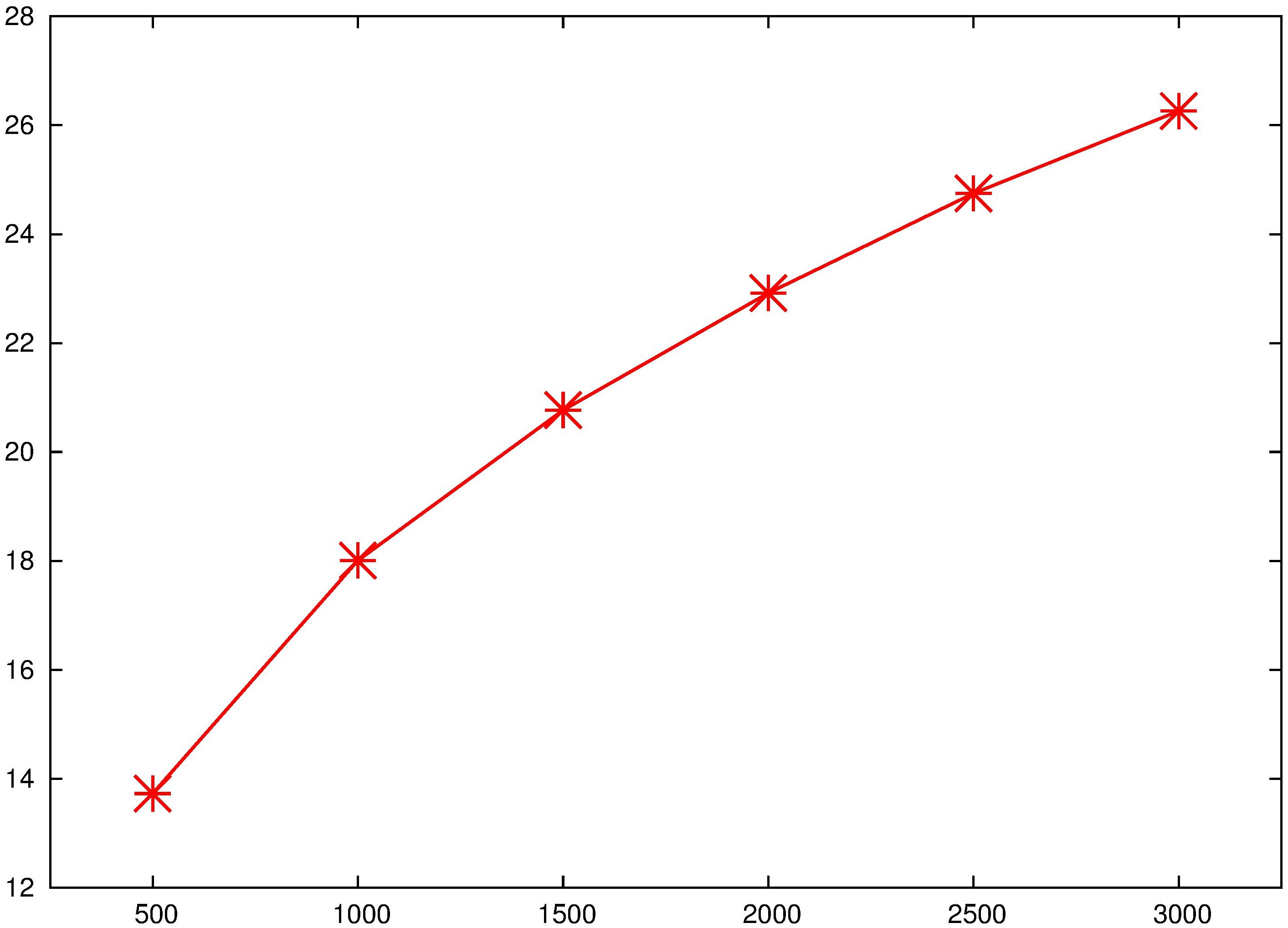}
  \caption{Streamwise coordinate (after the beginning of the step) of the reattachment point for the base flow (in ordinate),
   as a function of the Reynolds number (in abscissa).}
  \label{sepa}
 \end{figure}

\begin{figure}[b]
  \centering
  \includegraphics[scale=0.25]{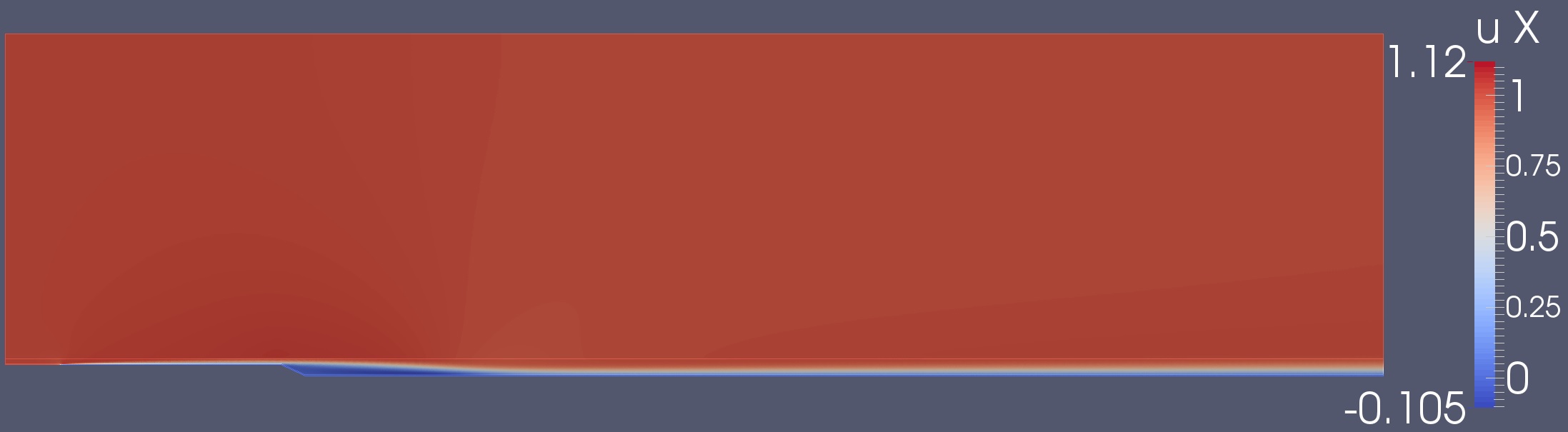}
  \caption{Horizontal component $U$ of the base-flow velocity at $\mathrm{Re}=1000$.}
  \label{cb1000}
 \end{figure}

 We have taken into consideration Reynolds numbers ranging from 500 to 3000, with increments of 500.
 The flow separates from the bottom boundary at the step and, with growing Reynolds number, 
 a larger and larger recirculation bubble develops in the wake, until reattachment takes place.
 Figure~\ref{sepa} displays the dependence of the reattachment point (i.e.\ the abscissa, after the beginning of the step,
 at which the vertical derivative of the horizontal velocity at the lower wall turns from negative to positive) as a function of $\mathrm{Re}$.
 A sketch of the base flow for $\mathrm{Re}=1000$ is presented in figure~\ref{cb1000}.

 \section{Linear stability analysis} \label{lsa}

 After fixing our base flow, we perform a linear stability analysis (see e.g.\ \cite{MPS12,V07}).
 Owing to the steadiness of the base flow and to its invariance in the spanwise direction,
 we can expand the perturbation field from (\ref{dec}) onto an exponential basis in both $t$ and $z$:
 \begin{equation*}
  \binom{\bm{u}'}{p'}\!(\bm{x},t)=\binom{\bm{\mathcal{U}}}{\mathcal{P}}\!(\bm{x})\,\ue^{\sigma t}+\textrm{c.c.}=\binom{\bm{u}''}{p''}\!(x,y)\,\ue^{\ui\beta z+\sigma t}+\textrm{c.c.}\qquad\beta\in\mathbb{R},\ \sigma\in\mathbb{C}\;.
 \end{equation*}
 The resulting linearized equation for the (direct) perturbation is:
 \begin{equation} \label{dirper}
  \left\{\begin{array}{rcl}\sigma\bm{\uu}+\bm{\uu}\cdot\bm{\nabla}\bm{U}+\bm{U}\cdot\bm{\nabla}\bm{\uu}\!\!&\!\!=\!\!&\!\!-\bm{\nabla}\mathcal{P}+\nu\nabla^2\bm{\uu}\;,\\
  \bm{\nabla}\cdot\bm{\uu}\!\!&\!\!=\!\!&\!\!0\;.\end{array}\right.
 \end{equation}
 \begin{figure}[t]
  \centering
  \includegraphics[scale=0.15]{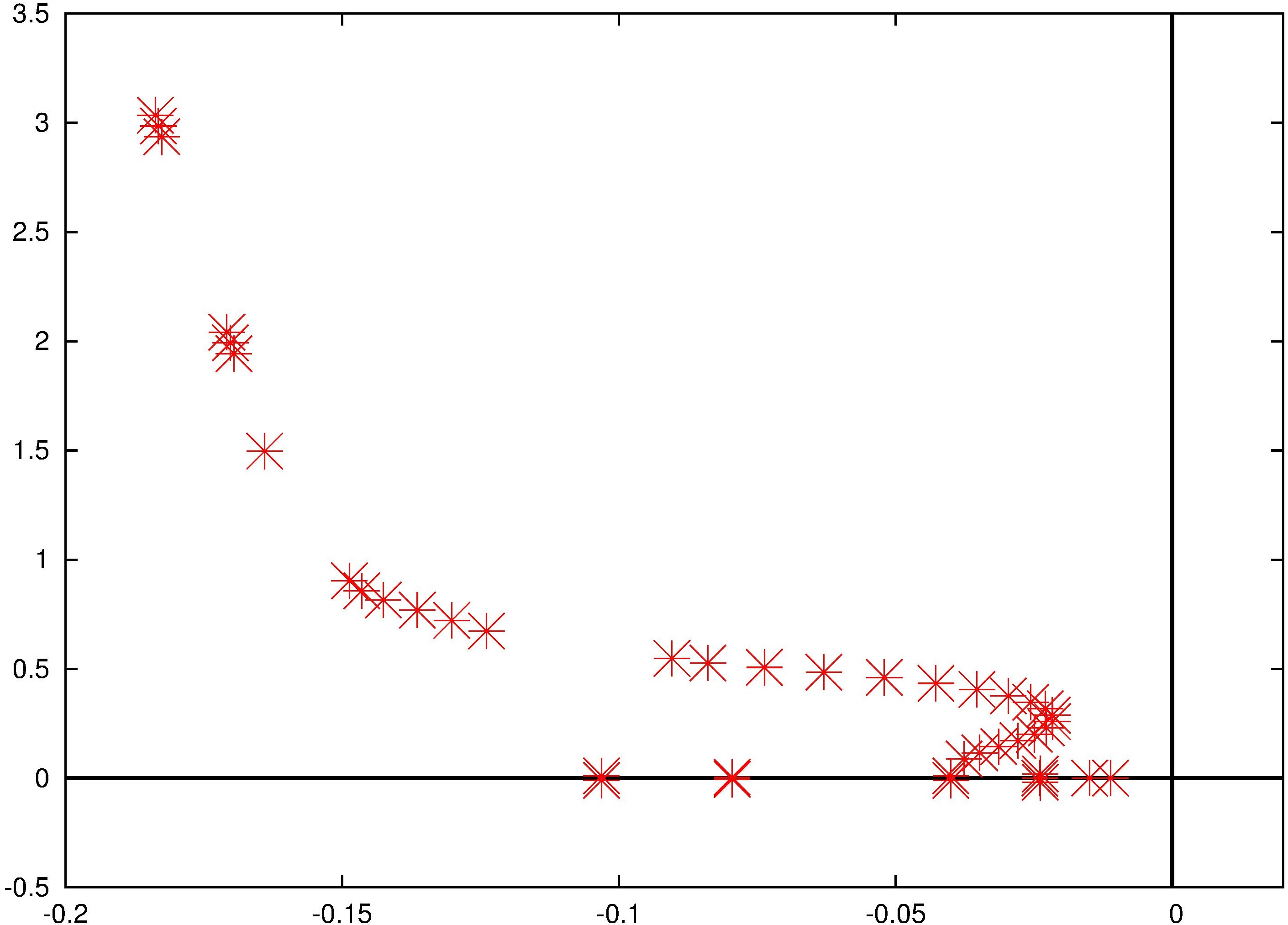}
  \caption{Complex spectrum of (\ref{dirper}) for the perturbation field, at $\mathrm{Re}=1000$ and $\beta=0$. The real and imaginary parts of $\sigma$
   are plotted in abscissa and ordinate, respectively.}
  \label{spe1000_0}
 \end{figure}%
 \begin{figure}[b!]
  \centering
  \includegraphics[scale=0.15]{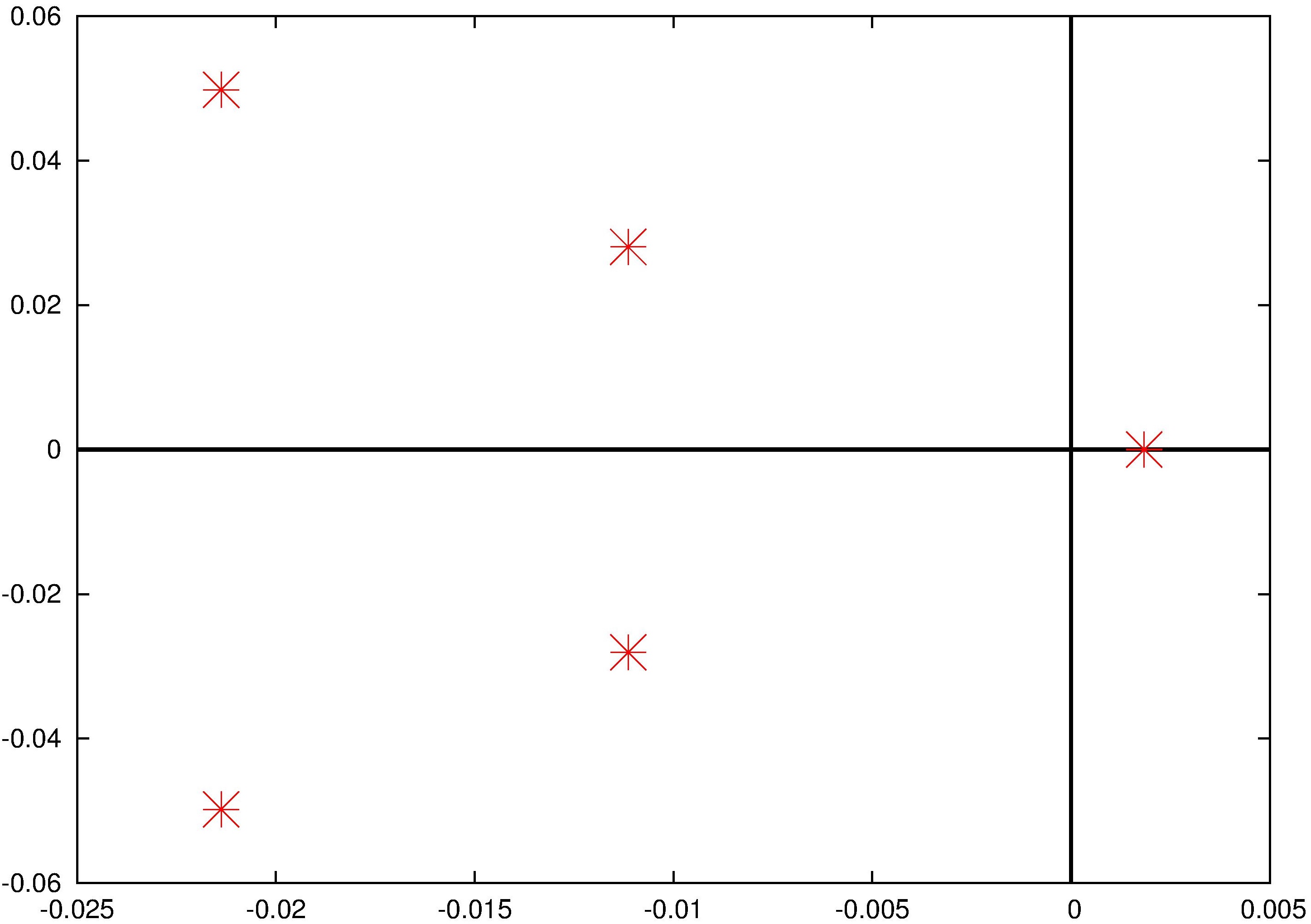}
  \caption{Same as in figure~\ref{spe1000_0} but for $\beta=1$.}
  \label{spe1000_1}
 \end{figure}%
 \begin{figure}[t]
  \centering
  \includegraphics[scale=0.15]{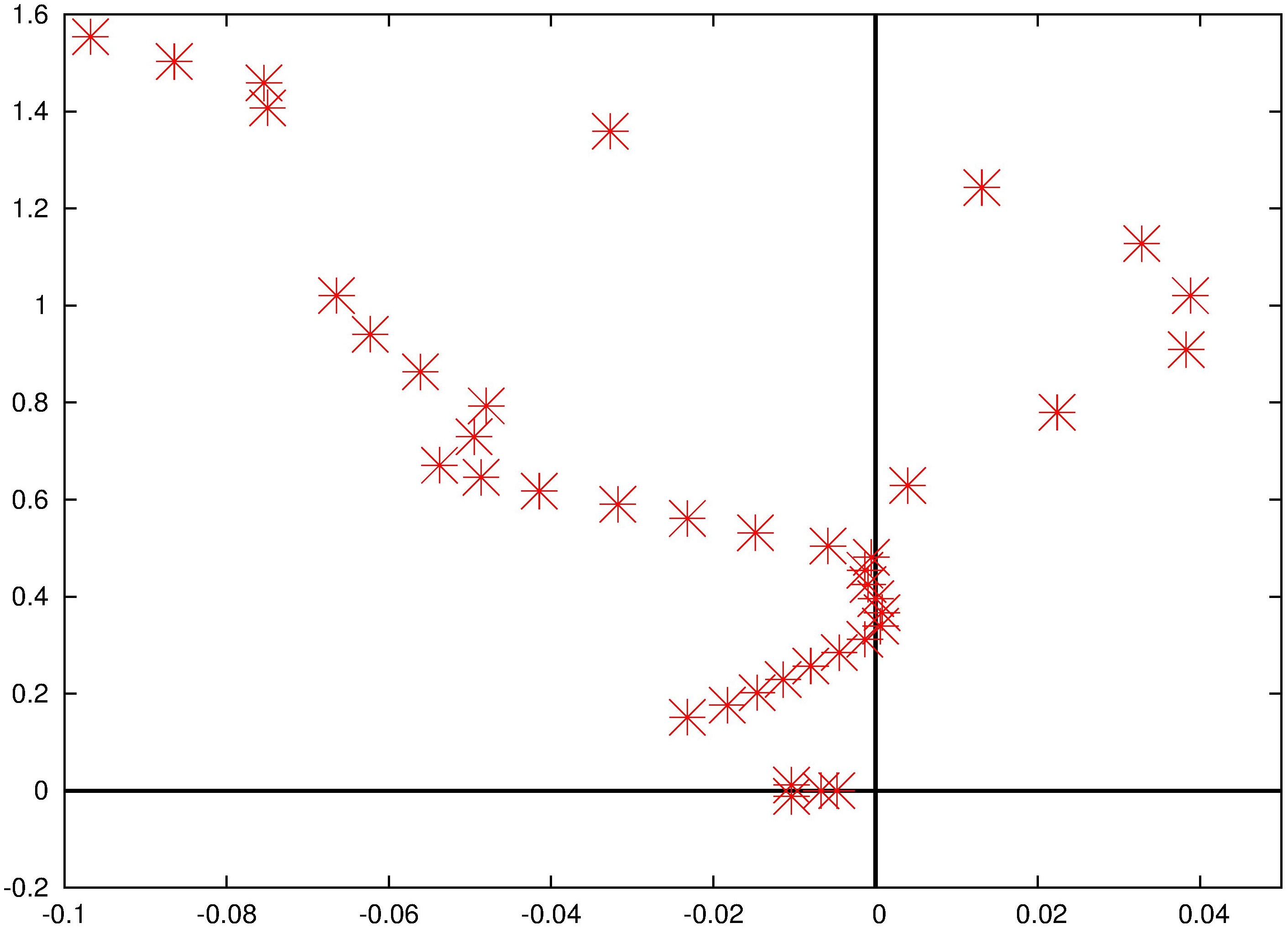}
  \caption{Same as in figure~\ref{spe1000_0} but for $\mathrm{Re}=3000$.}
  \label{spe3000_0}
 \end{figure}%
 We then look for the spectrum of the eigenvalues, i.e.\ complex values of $\sigma$ such that (\ref{dirper}) has nontrivial solutions
 $\displaystyle\binom{\bm{\uu}}{\mathcal{P}}$, which can accordingly be defined as (direct) eigenvectors.
 The baseline case, $\mathrm{Re}=1000$ and $\beta=0$ (i.e.\ no spanwise dependence), is plotted in figure~\ref{spe1000_0},
 and is stable because all the eigenvalues have negative real part. Instability can be reached in two ways: either modifying
 the spanwise wavenumber (e.g.\ $\beta=1$ in figure~\ref{spe1000_1}), or by augmenting the Reynolds number (e.g.\ $\mathrm{Re}=3000$
 in figure~\ref{spe3000_0}). It is worth noticing that the former operation induces stationary instabilities --- as the imaginary part
 of the rightmost eigenvalue still vanishes --- while the latter introduces unstationary ones ($\Im(\sigma)\neq0$ for those points where
 $\Re(\sigma)>0$, and the picture is of course symmetric with respect to the horizontal axis). The critical Reynolds number, for which
 the flow develops its first linear instability at some value of $\beta$, is approximately 750.\\
 A sketch of the perturbation field (for the largest-real-part eigenvalue depicted in figure~\ref{spe1000_1}),
 at $\mathrm{Re}=1000$ and $\beta=1$, is presented in figure~\ref{dir1000_1}.
 It is evident that this eigenvector is a physically meaningful one because it is concentrated in the recirculation bubble,
 which is the zone where instabilty develops.
 \begin{figure}[b!]
  \centering
  \includegraphics[scale=0.25]{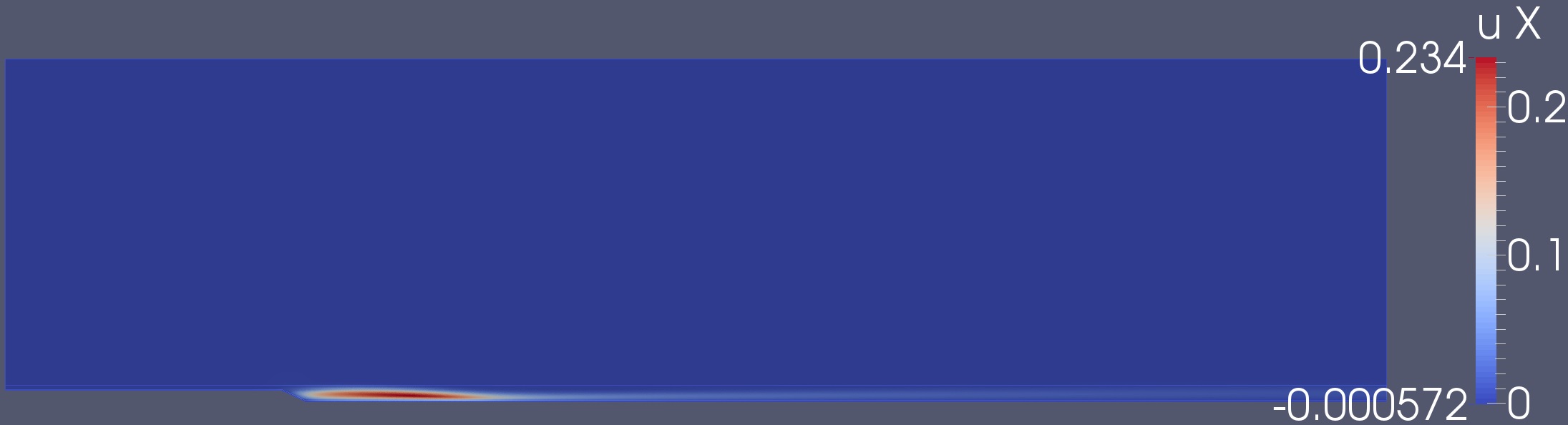}
  \caption{Horizontal component $u''$ of the dominant perturbation field at $\mathrm{Re}=1000$ and $\beta=1$.}
  \label{dir1000_1}
 \end{figure}%

 \section{Control and gain} \label{con}
 
 In this section we follow \cite{P09,M07} by introducing a forcing on the right-hand side of the Navier--Stokes equation,
 which we assume as steady and volumic: $\bm{\mathcal{F}}(\bm{x},\xcancel{t})=\bm{f}(x,y)\ue^{\ui\beta z}$.
 A key point is that this allows us not only to leave the boundary conditions unchanged with respect to (\ref{dirper}),
 but more importantly to confine ourselves to \emph{steady} solutions $\displaystyle\binom{\bm{\uu}}{\mathcal{P}}$. Indeed, we have already analyzed
 and found the temporal evolution of the general unforced solutions (eigenvalues and eigenvectors) in the previous section,
 and what we are looking for here is just a particular solution to a steady forcing, which can thus be assumed as time-independent.\\
 Therefore we focus on the equations:
 \begin{equation} \label{forper}
  \left\{\begin{array}{rcl}\bm{\uu}\cdot\bm{\nabla}\bm{U}+\bm{U}\cdot\bm{\nabla}\bm{\uu}\!\!&\!\!=\!\!&\!\!-\bm{\nabla}\mathcal{P}+\nu\nabla^2\bm{\uu}+\bm{\mathcal{F}}\;,\\
  \bm{\nabla}\cdot\bm{\uu}\!\!&\!\!=\!\!&\!\!0\;.\end{array}\right.
 \end{equation}
 We define as \emph{gain} the quantity
 \begin{equation*}
  g\equiv\frac{E_u}{E_f}=\frac{\int\ud x\int\ud y\,|\bm{u}''|^2}{\int\ud x\int\ud y\,|\bm{f}|^2}\;,
 \end{equation*}
 and as \emph{optimal} gain:
 \begin{equation*}
  G(\beta,\mathrm{Re})\equiv\max_{\bm{\mathcal{F}}}g\;.
 \end{equation*}
 The procedure to find the optimal forcing consists in an iterative algorithm making use of the \emph{adjoint} variables
 $\displaystyle\binom{\bm{\uu}^{\dag}}{\mathcal{P}^{\dag}}$ satisfying
 \begin{equation} \label{adjper}
  \left\{\begin{array}{rcl}(\bm{\nabla}\bm{U})\cdot\bm{\uu}^{\dag}-\bm{U}\cdot\bm{\nabla}\bm{\uu}^{\dag}\!\!&\!\!=\!\!&\!\!\displaystyle\bm{\nabla}\mathcal{P}^{\dag}+\nu\nabla^2\bm{\uu}^{\dag}+\frac{\bm{\uu}}{2E_f}\;,\\
  \bm{\nabla}\cdot\bm{\uu}^\dag\!\!&\!\!=\!\!&\!\!0\;,\end{array}\right.
 \end{equation}
 coupled with suitable outlet boundary conditions (see e.g.\ \cite{MBPG14}) specified in appendix~\ref{app}.
 
 \subsection{Standard (non-penalized) case}
 
\begin{figure}
  \centering
  \includegraphics[scale=0.3]{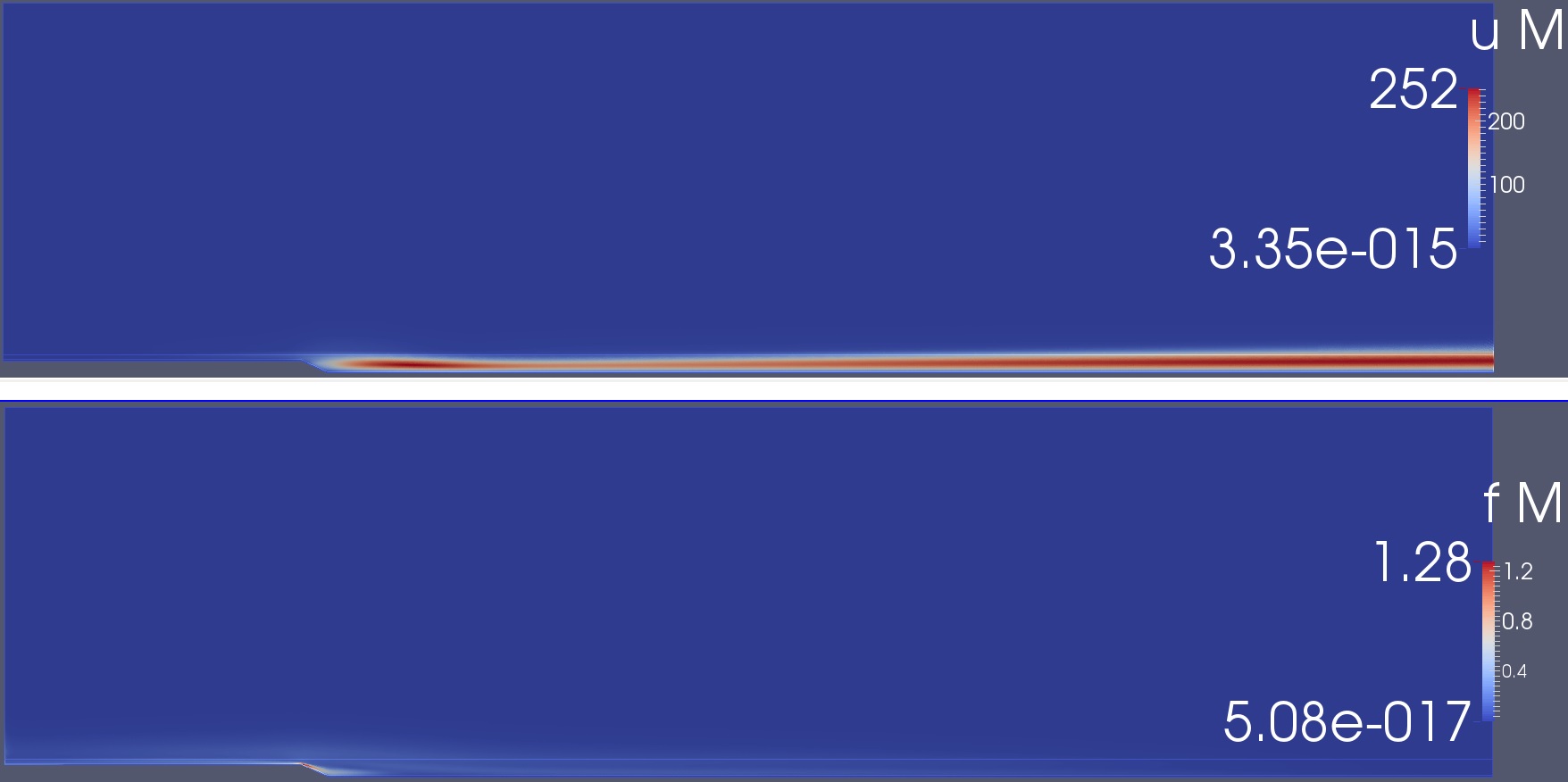}
  \caption{Magnitude of the optimal response and of the corresponding optimal forcing fields, $|\bm{u}''|$ and $|\bm{f}|$,
   in the upper and lower panels respectively, at $\mathrm{Re}=1000$ and $\beta=1$. The gain is maximized according to scheme (\ref{forper}).}
  \label{nonpen}
 \end{figure}

 In our reference case (standard geometry with $\mathrm{Re}=1000$ and $\beta=1$) the magnitudes of what we have obtained numerically
 as optimal response and corresponding optimal forcing are sketched in figure~\ref{nonpen}.\\
 A clear problem arises here: even if on the one hand the optimal response develops streamwise for the whole length,
 on the other hand the optimal forcing is localized in correspondence of the step and of the sloping portion of the wall.
 This is not what one would expect for physical realizability, as on the contrary a localization on the horizontal upstream part
 would be suitable. Indeed, our aim is to take advantage of the streak lift-up, i.e.\ the formation of counter-rotating longitudinal vortices
 (see e.g.\ \cite{CPD09,DCP14a,DCP14b}),
 and to let these latter interact with the recirculation bubble, in an interesting example of interaction between the Kelvin--Helmholtz and the wake
 instabilities. In \cite{PCD10,PDC10,PDC11} the formation of streaks was implemented experimentally by placing a series of small cylinders,
 acting as rugosity elements.%
 \footnote{The modified flow that one would obtain after physically placing the roughness element is clearly not the same as the one in their absence.
  What is meant here is that this discrepancy must be small for our theory to work (which clearly poses severe restrictions on the applicable elements),
  so that the modified flow should be obtainable as the sum of the original basic flow plus the weak perturbations currently analyzed.}
 (Notice that the spanwise periodicity of this array of cylinders can be described effectively via our
 periodic expansion in $z$, i.e.\ by means of the wavenumber $\beta$ which should equal $2\pi$ divided by the array spacing.)
 In principle, one could introduce step functions in the integrals defining the gain, and we briefly explored this option preliminarily.
 However, in the next subsection we are going to attack this problem by means of a \emph{penalization} method (see e.g.\ \cite{MPS08,MBPS11}).

 \subsection{Penalized case}

 Let us momentarily introduce an effective-viscosity field, $\nu_{\mathrm{eff}}(x)$, defined as equal to $\nu$ upstream and until
 the beginning of the step, and to a value some orders of magnitude larger for abscissae downstream of it. We then focus on:
 \begin{equation} \label{penper}
  \left\{\begin{array}{rcl}\bm{\uu}\cdot\bm{\nabla}\bm{U}+\bm{U}\cdot\bm{\nabla}\bm{\uu}\!\!&\!\!=\!\!&\!\!-\bm{\nabla}\mathcal{P}+\nu_{\mathrm{eff}}(x)\nabla^2\bm{\uu}+\bm{\mathcal{F}}\;,\\
  \bm{\nabla}\cdot\bm{\uu}\!\!&\!\!=\!\!&\!\!0\;.\end{array}\right.
 \end{equation}
 \begin{figure}[t]
  \centering
  \includegraphics[scale=0.3]{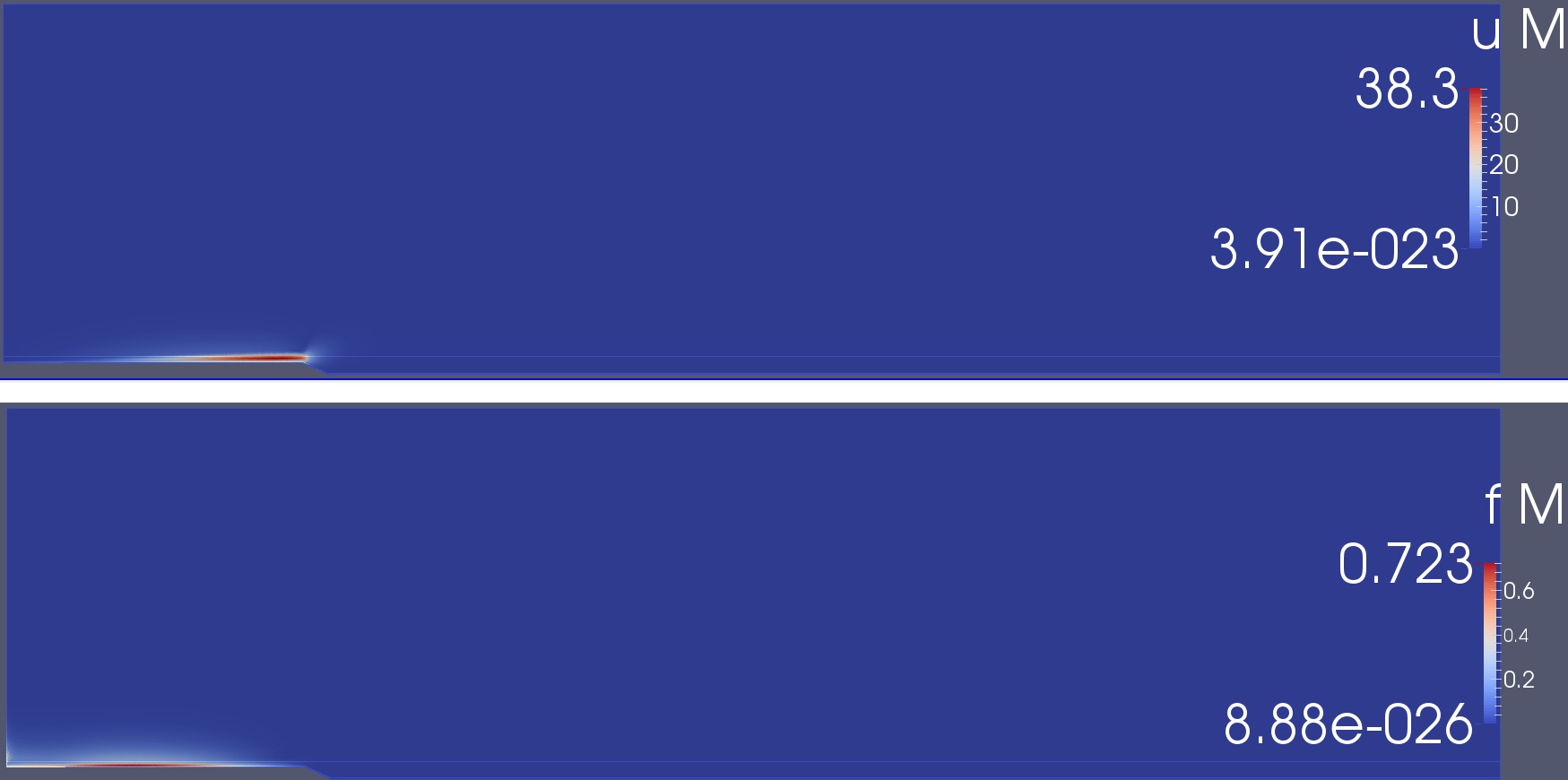}
  \caption{Magnitude of the sub-optimal (penalized) response and forcing fields, $|\bm{u}''|$ and $|\bm{f}|$, in the upper and lower panels respectively,
   at $\mathrm{Re}=1000$ and $\beta=1$. The gain is maximized following (\ref{penper}).}
  \label{pen}
 \end{figure}%
 In this way we obtain the optimal response and forcing sketched in figure~\ref{pen}, which should more precisely be defined as
 \emph{sub-optimal} because of the penalization scheme. The shape of the forcing is now what we expect for physical realizability,
 due to its localization on the upstream portion of the wall, but the same cannot be said about the velocity response, due to its concentrated character
 very different in look from the envisaged streaks appearing in the previous subsection.\\
  \begin{figure}[b!]
  \centering
  \includegraphics[scale=0.15]{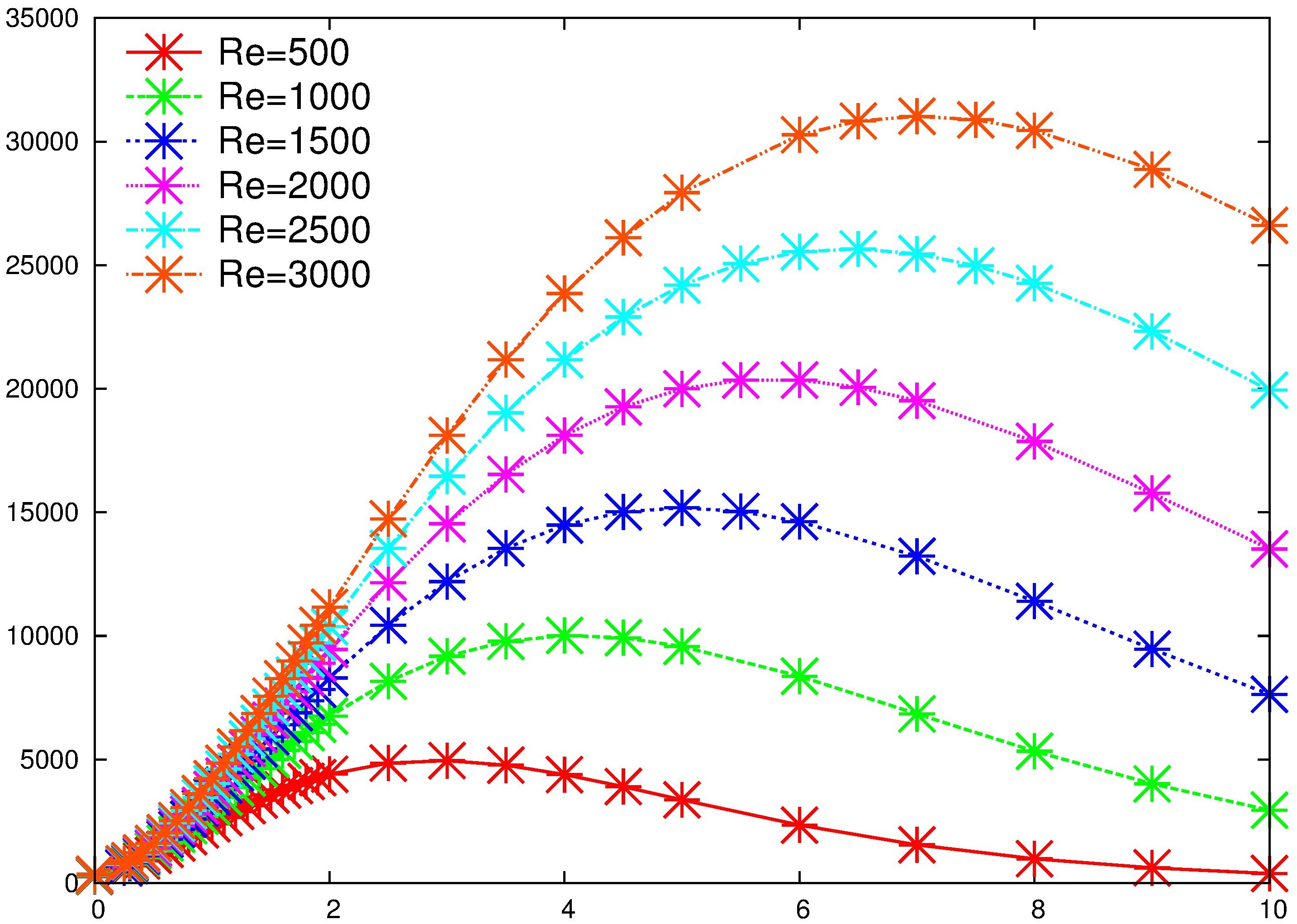}
  \caption{Optimal gain (in ordinates) vs.\ spanwise wavenumber (in abscissa) at different Reynolds numbers.
   The gain is maximized according to scheme (\ref{penper}).}
  \label{pena}
 \end{figure}%
 In figure~\ref{pena} we plot the optimal gain $G$ as a function of $\beta$ at different $\mathrm{Re}$.
 The maximum of the curve not only obviously grows with $\mathrm{Re}$, but also shifts to the right.
 As on the contrary the boundary-layer thickness shrinks when increasing the Reynolds number, it is interesting to plot
 the product between the generalized displacement computed at the step, $\delta_1|_{x=0}$, and the optimal wavenumber $\beta_{\mathrm{opt}}$.
 This is done in figure~\ref{betdel} and proves that the ratio between the thickness and the optimal spanwise wavelength is almost independent of
 $\mathrm{Re}$.
  \begin{figure}
  \centering
  \includegraphics[scale=0.15]{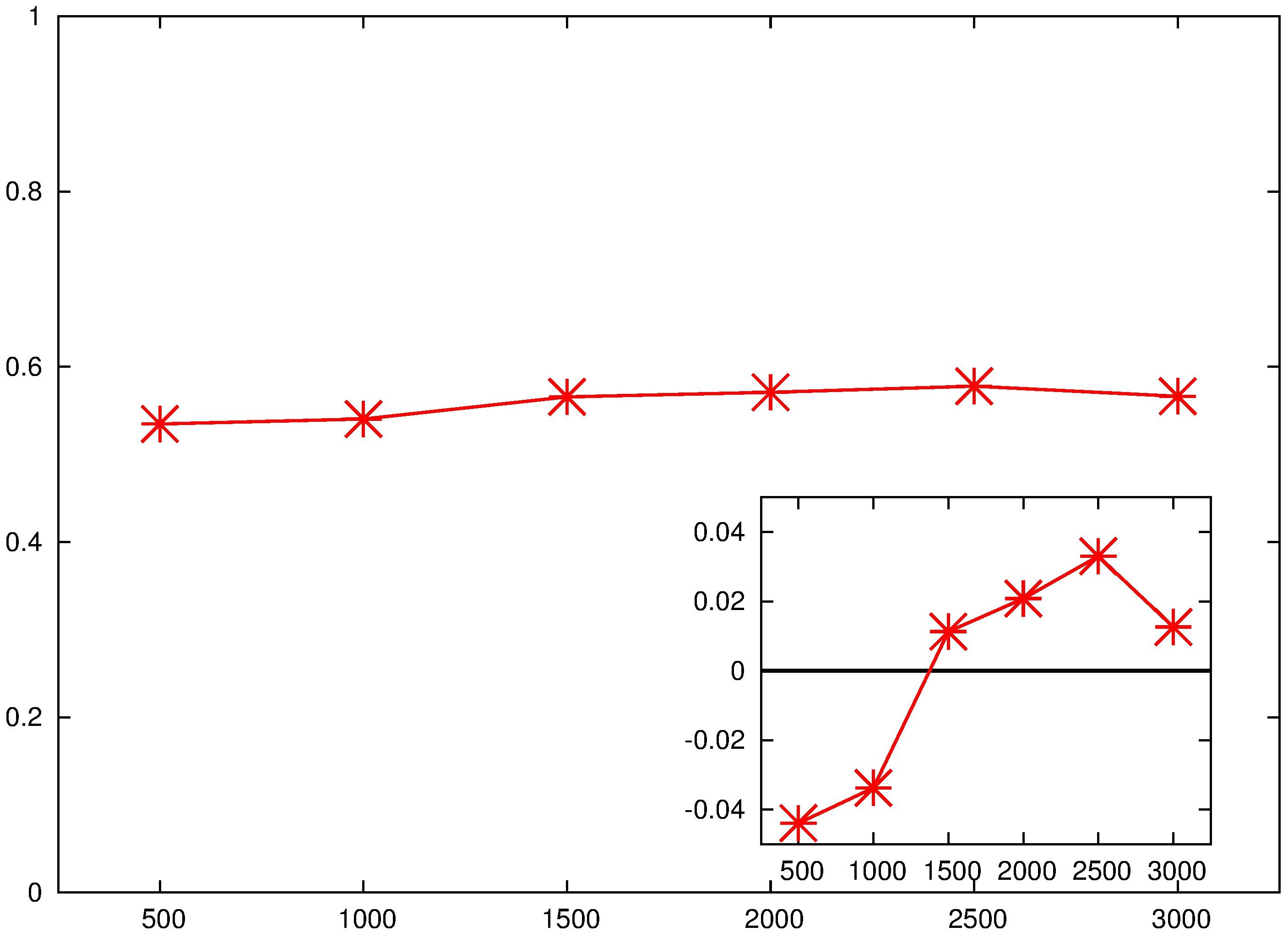}
  \caption{Product between optimal spanwise wavenumber --- maxima of figure~\ref{pena} --- and generalized displacement thickness at the step
   $\delta_1|_{x=0}$ (in ordinate), vs.\ Reynolds number (in abscissa).
   The inset shows that, within a relative maximum error of less than 5\%, this product is independent of $\mathrm{Re}$.}
  \label{betdel}
 \end{figure}%

 \subsection{Penalized control with non-penalized response}

 The way to circumvent the paradox presented in the previous subsection is very easy.
 One can indeed find the optimal control through the penalized scheme, but of course once this forcing has been found
 its real action on the physical velocity must be computed with the actual (space-independent) viscosity $\nu$.
 We then implement what one could call ``non-penalized response to penalized-optimal control'':
 \begin{equation} \label{pnpper}
  \begin{array}{r}\phantom{a}\\\left.\!\begin{array}{r}\textrm{LAST}\\\textrm{STEP}\end{array}\!\!\right\{\!\end{array}\!\!\!\begin{array}{rcl}\bm{\uu}\cdot\bm{\nabla}\bm{U}+\bm{U}\cdot\bm{\nabla}\bm{\uu}\!\!&\!\!=\!\!&\!\!-\bm{\nabla}\mathcal{P}+\nu_{\mathrm{eff}}(x)\nabla^2\bm{\uu}+\bm{\mathcal{F}}\;,\\
  \bm{\nabla}\cdot\bm{\uu}\!\!&\!\!=\!\!&\!\!0\;,\\
  \bm{\uu}\cdot\bm{\nabla}\bm{U}+\bm{U}\cdot\bm{\nabla}\bm{\uu}\!\!&\!\!=\!\!&\!\!-\bm{\nabla}\mathcal{P}+\nu\nabla^2\bm{\uu}+\bm{\mathcal{F}}_{\mathrm{opt}}\;.\end{array}\!\!\!\begin{array}{l}\!\left\}\!\!\begin{array}{l}\textrm{LOOP TO}\\\textrm{FIND }\bm{\mathcal{F}}_{\mathrm{opt}}\end{array}\!\right.\\\phantom{a}\end{array}
 \end{equation}
 \begin{figure}
  \centering
  \includegraphics[scale=0.3]{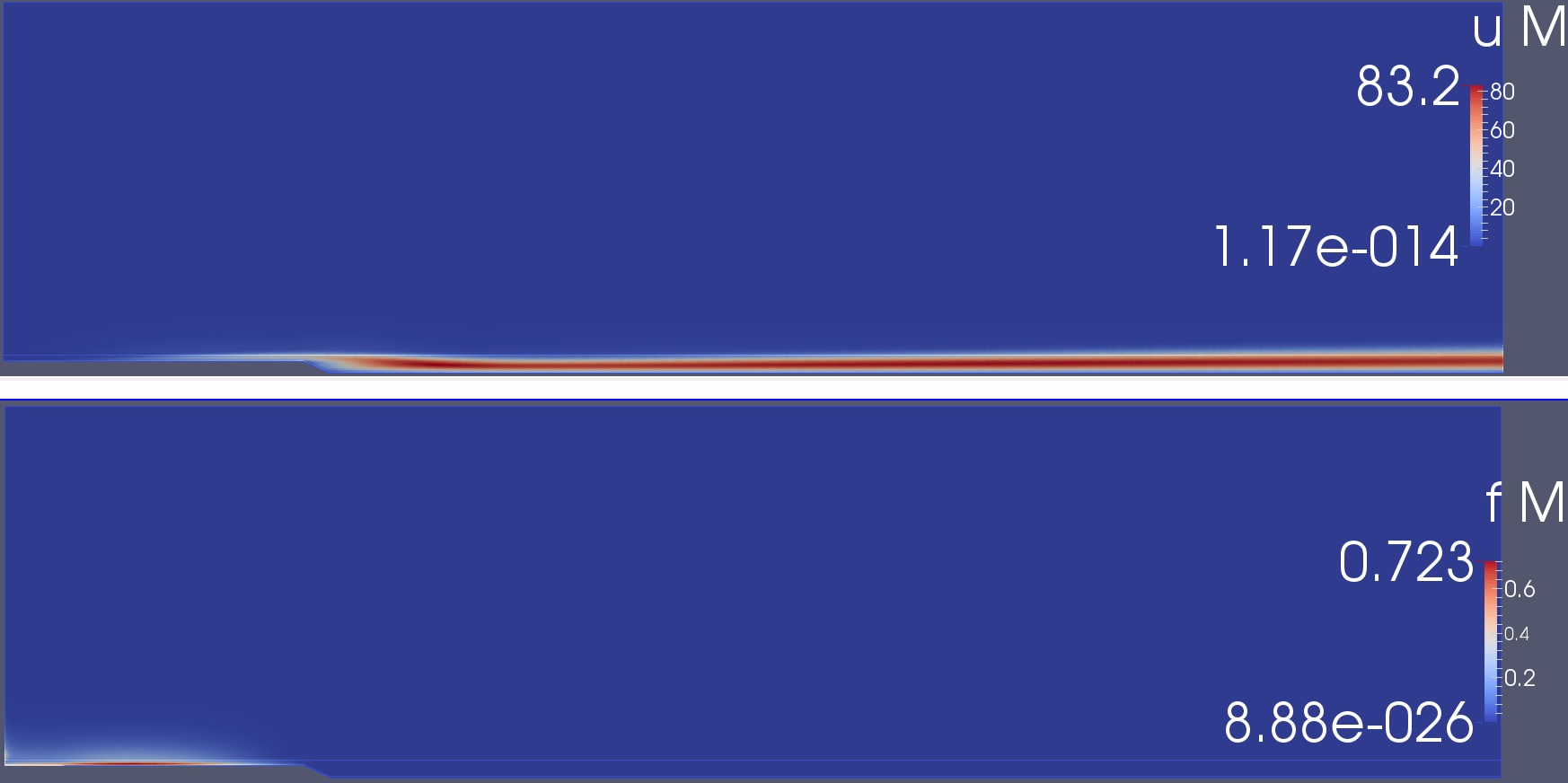}
  \caption{Magnitude of the non-penalized response to penalized-optimal control, $|\bm{u}''|$, in the upper panel, according to the scheme (\ref{pnpper}),
   at $\mathrm{Re}=1000$ and $\beta=1$. The gain is maximized according to (\ref{penper}) and the magnitude of the corresponding penalized-optimal forcing
   field $|\bm{f}|$ (the same as in figure~\ref{pen}) is reported again in the lower panel for the sake of simplicity.}
  \label{pennonpen}
 \end{figure}%
 In this way we obtain the response sketched in figure~\ref{pennonpen}, together with the aforementioned penalized-optimal forcing field.
 The fact that streaks are actually generated is confirmed by figure~\ref{g18}, which represents vertical cuts of the domain
 $y\in[0,1]\times z\in[0,2\pi)$ at eight different streamwise locations.\\
 \begin{figure}[p]
  \centering
  \includegraphics[scale=0.11]{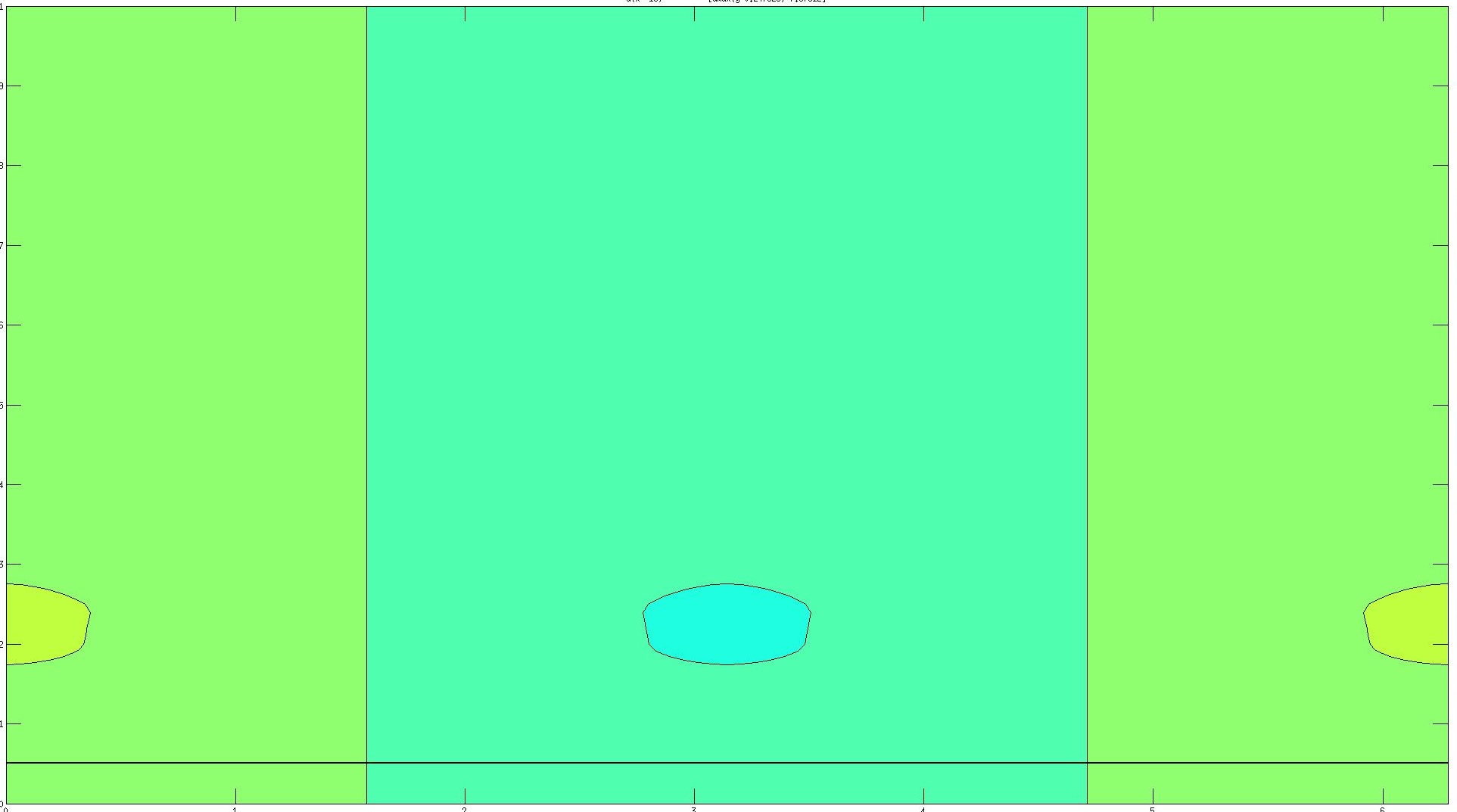}\hfill
  \includegraphics[scale=0.11]{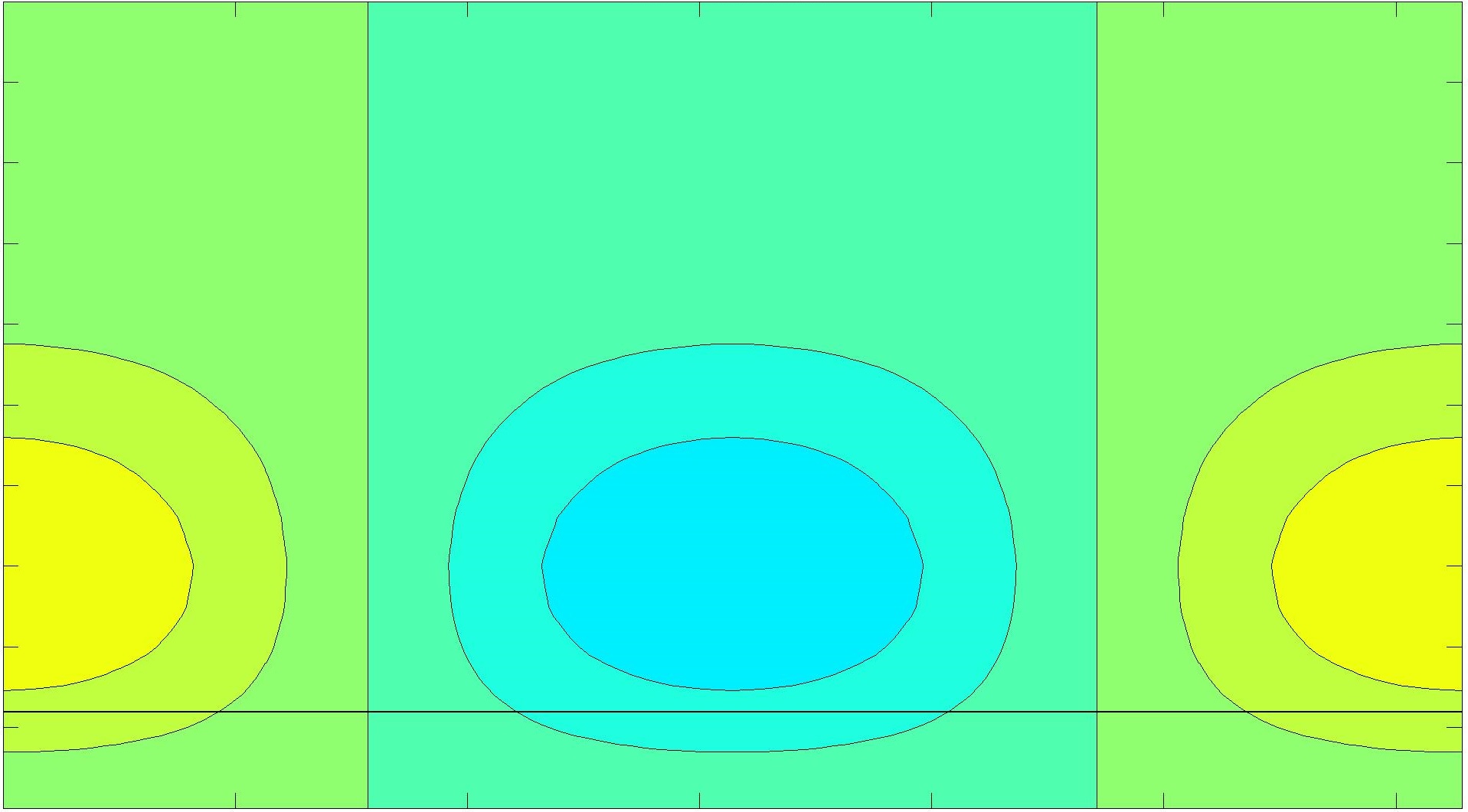}\\[0.6cm]
  \includegraphics[scale=0.11]{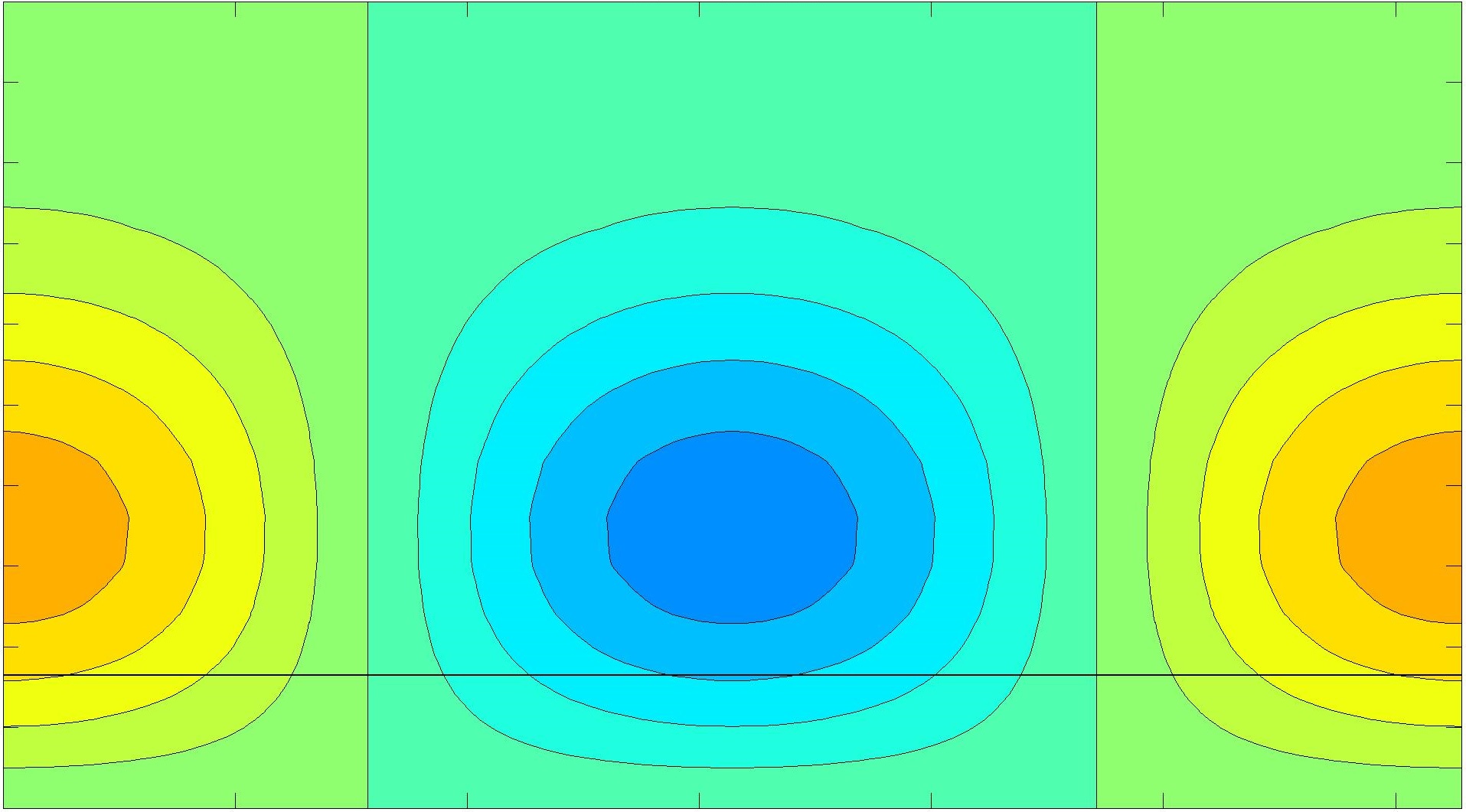}\hfill
  \includegraphics[scale=0.11]{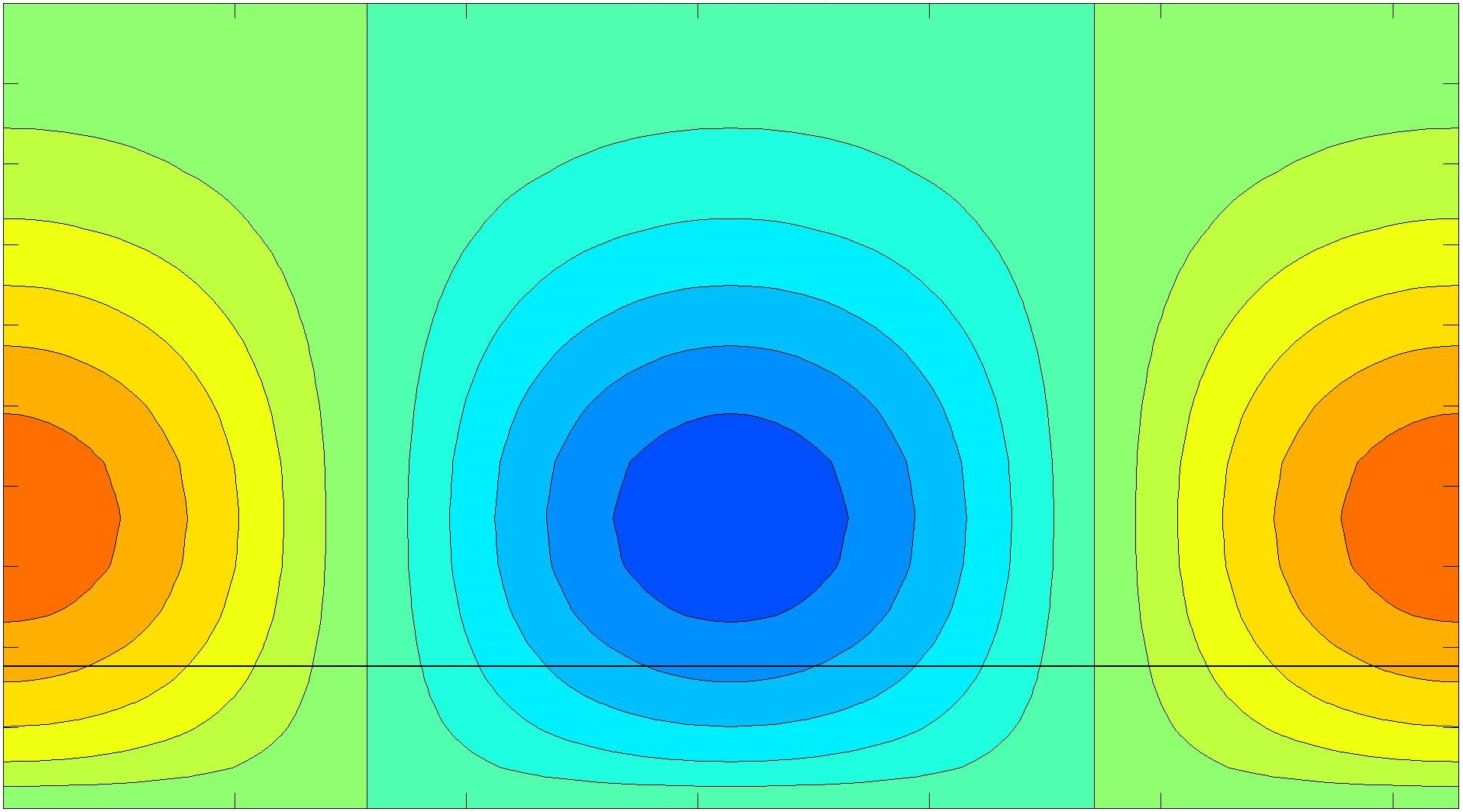}\\[0.6cm]
  \includegraphics[scale=0.11]{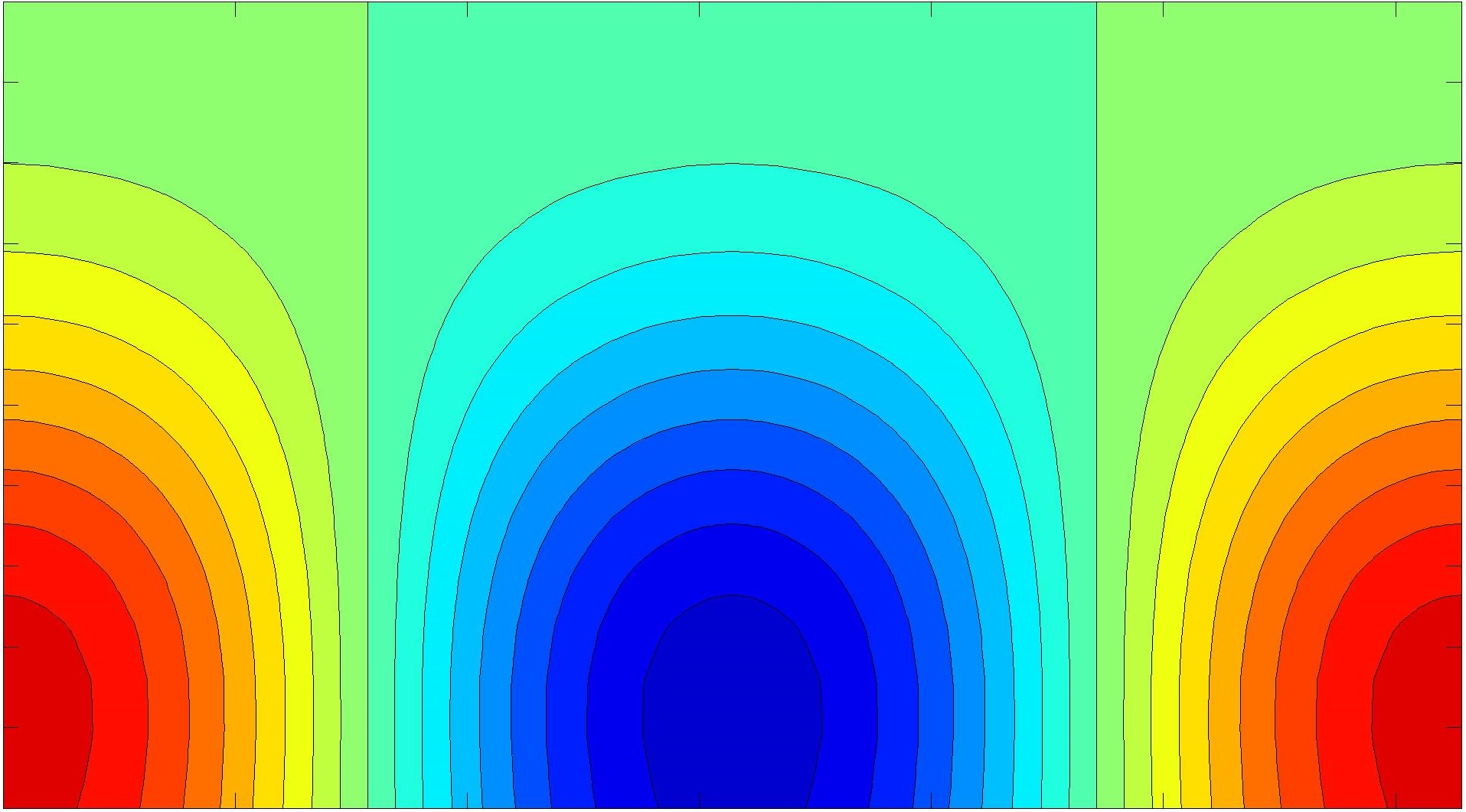}\hfill
  \includegraphics[scale=0.11]{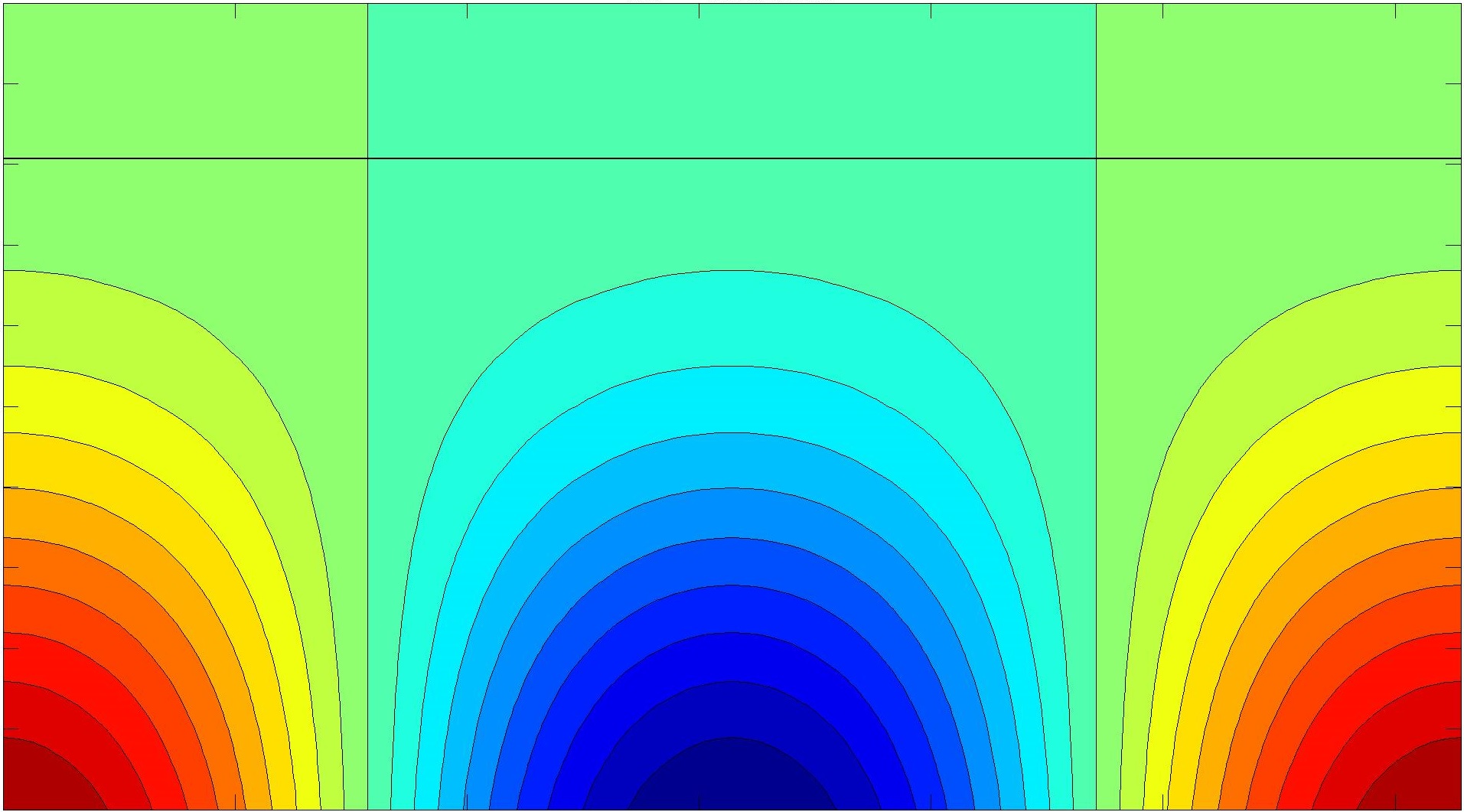}\\[0.6cm]
  \includegraphics[scale=0.11]{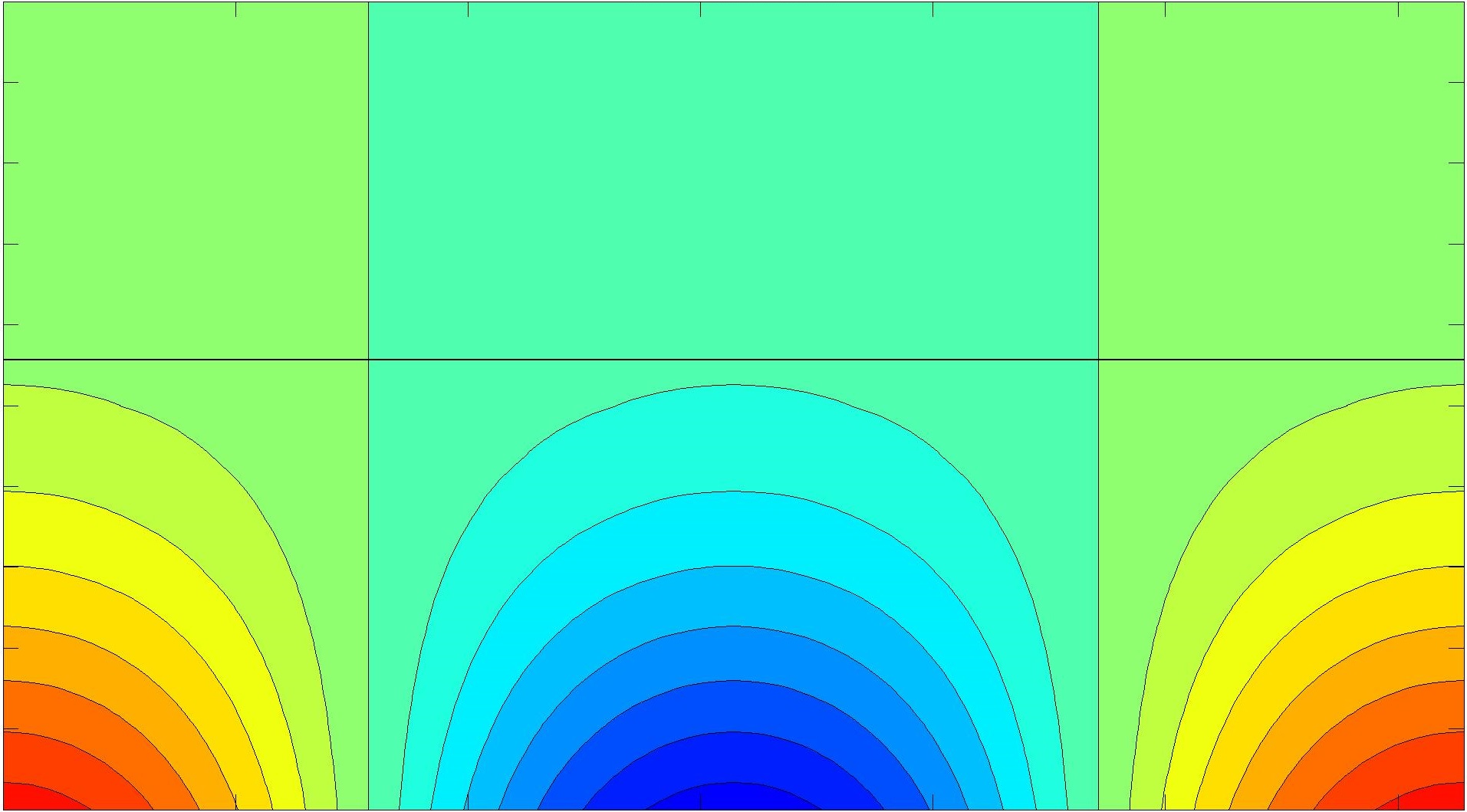}\hfill
  \includegraphics[scale=0.11]{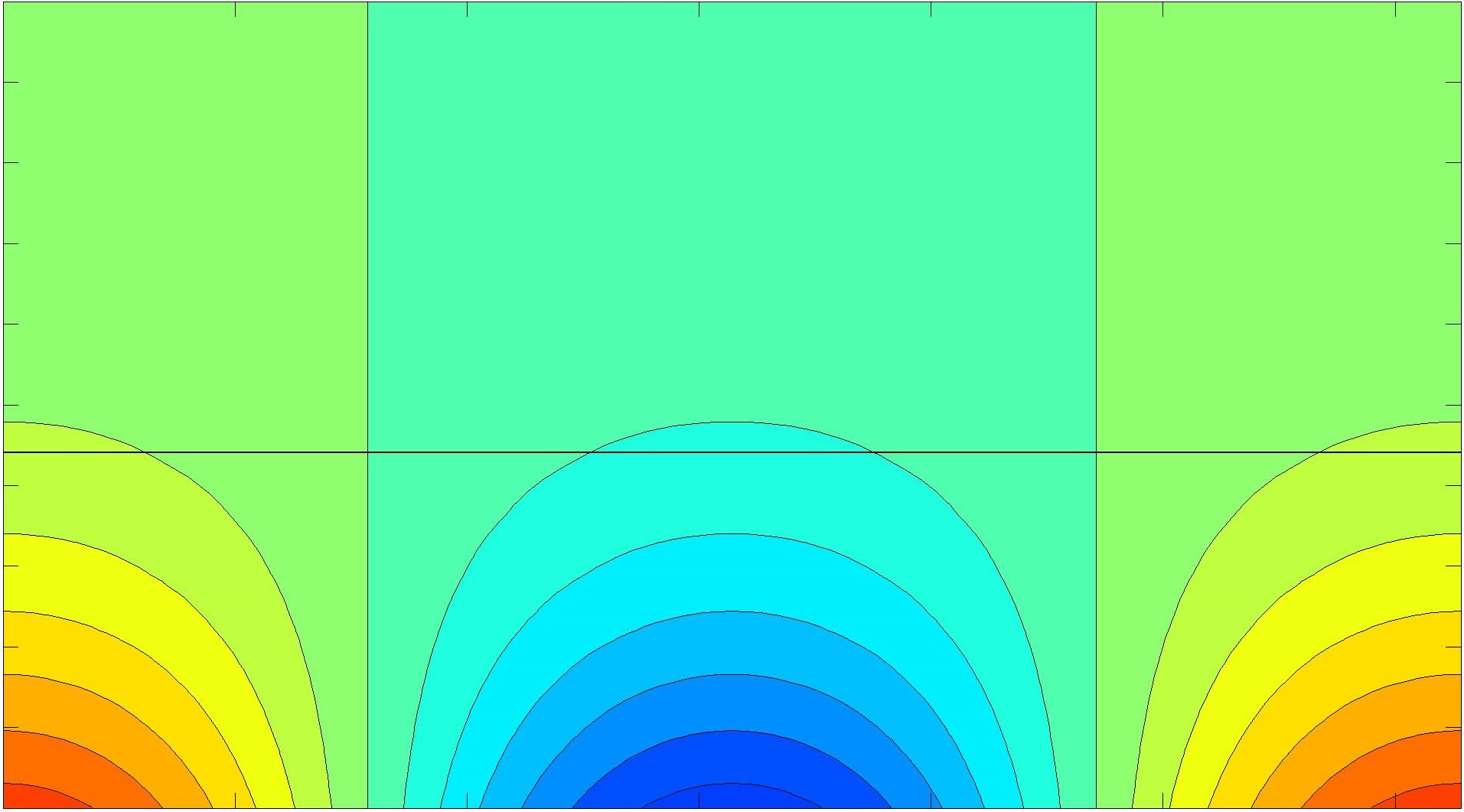}\\
  \caption{Streamwise component $u''$ of the optimal response (velocity field depicted in the upper panel of figure~\ref{pennonpen}),
   with positive values in red and negative ones in blue, in vertical cuts at eight different streamwise coordinates:
   $x=-15$ and $-10$ (top row), $-5$ and $0$, $5$ and $10$, $15$ and $20$ (bottom row).
   The horizontal axis is $z\in[0,2\pi)$ and the vertical one is $y\in[0,1]$; notice that for the four latter plots
   the physical domain extends below the bottom border of the figure, namely for a depth $-1\le y\le0$ which exactly equals the height shown.
   The black horizontal lines represent the height of the generalized displacement thickness $\delta_1(x)$ at each location;
   the line is not shown in the fifth panel because happening to be placed above the top border, i.e.\ $\delta_1|_{x=5}>1$.}
  \label{g18}
 \end{figure}%
 In figure~\ref{tutt} we plot a comparison for the optimal $G$ as a function of $\beta$ at $\mathrm{Re}=500$ according to the three aforementioned schemes.
 In this completely-stable situation, one can see that --- moving from the initial non-penalized scheme (black) to the final scheme proposed in this
 subsection (blue) --- the loss in the optimal gain consists in less than one order of magnitude, but with the advantage of delivering an optimal forcing
 definitely feasible in terms of physical realizability.\\
  \begin{figure}[t]
  \centering
  \includegraphics[scale=0.15]{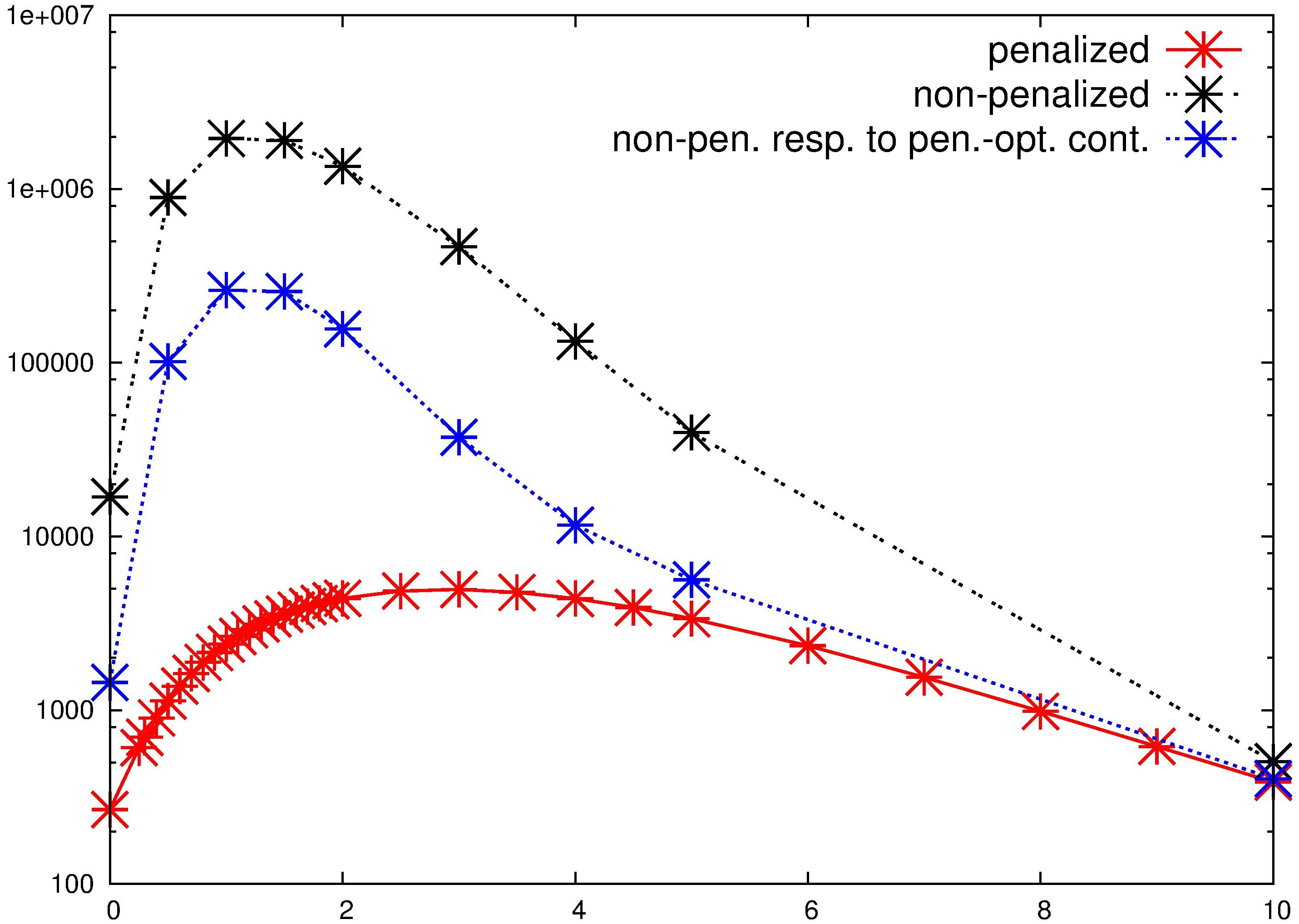}
  \caption{Optimal gain (in ordinates) vs.\ spanwise wavenumber (in abscissa) at $\mathrm{Re}=500$, according to the three maximization schemes
   (\ref{forper}) (black), (\ref{penper}) (red) and (\ref{pnpper}) (blue).}
  \label{tutt}
 \end{figure}%
 The comparison between (\ref{penper}) and (\ref{pnpper}) is also plotted in figure~\ref{ogn} at $\mathrm{Re}=3000$.
 Notice that scheme (\ref{forper}) cannot be enforced here, because the Reynolds number is larger than the critical value. 
 Also, since this configuration is unstable for some perturbations, we also plot (in black) the gain corresponding to a non-modal response field which is
 computed through an orthogonalization procedure, in order to exclude spurious peaks related to the modal amplification (not interesting here).
 \begin{figure}[b!]
  \centering
  \includegraphics[scale=0.15]{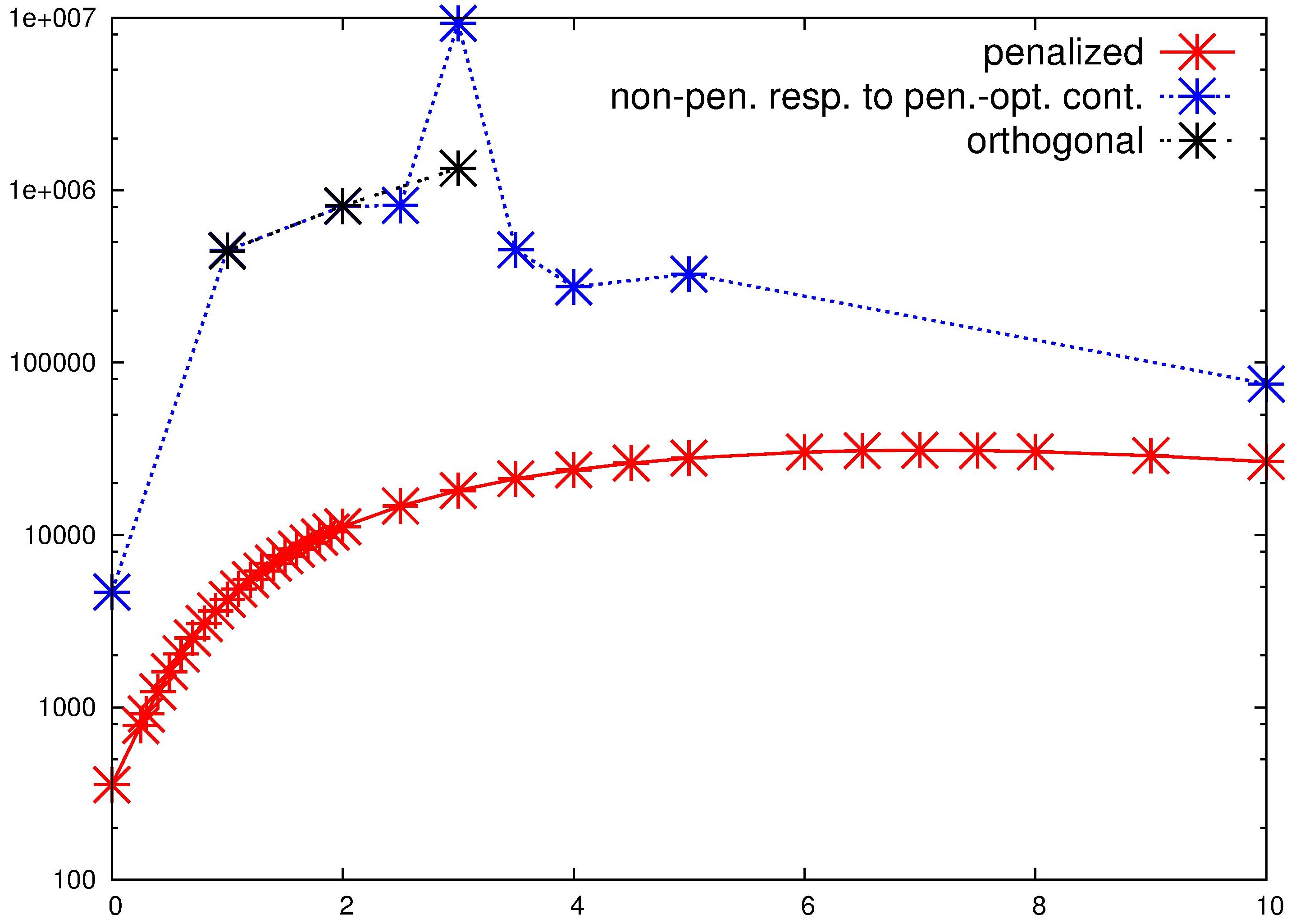}
  \caption{Optimal gain (in ordinates) vs.\ spanwise wavenumber (in abscissa) at $\mathrm{Re}=3000$, according to the two maximization schemes
   (\ref{penper}) (red) and (\ref{pnpper}) (blue). The black points are the result of a process of orthogonalization aimed at excluding spurious
   amplifications, as happening for the blue peak at $\beta=3$.}
  \label{ogn}
 \end{figure}%

 \section{Conclusions and perspectives} \label{per}
 
 We have studied the transient growth of perturbations in a separated boundary layer, namely in the wake of a backward-slanted step (at $25^{\circ}$).
 We have shown that, by means of a suitable penalization method, also unstable cases are tractable in our formalism.
 We have been able to find situations where the optimal control is spanwise-periodic and localized on the horizontal upstream portion of the wall
 (which experimentally can be reproduced by an array of rugosity elements), and the corresponding response is represented by streaks.
 
 Among future perspectives, five of them come to our mind.\\
 First, one could change the nature of the forcing term, from volumic to parietal. This would represent an external blowing
 or suction on the lower wall before the step, which would be implemented by imposing on this piece of boundary a condition
 on the velocity, that should keep zero tangential component but assume a certain nonzero normal component (a function of $x$).
 We have already made a preliminary test for this situation, but a more profound investigation is definitely required.\\
 Second, it would be interesting to relax the assumption of steady forcing and to investigate the problem also in the temporal domain.
 As a consequence, one should fix a finite time horizon for the optimization, and perform back-and-forth temporal loops until convergence \cite{TRHG06}.
 This is due to the fact that, by keeping the time dependence in the equations, the evolution of the adjoint field corresponds to a
 well-known backward-in-time integration, with ``final'' conditions imposed on the final time horizon. Of course the evolution
 of the direct field is forward-in-time, and one has to perform an optimization on the initial conditions.\\
 Also, the incompressibility of the flow is a key ingredient for the results shown here. It might be worth investigating
 how they change if a compressible flow is considered instead. We expect the theoretical analysis to be much more difficult,
 in view of the necessity of introducing an equation of state as well.\\
 Moreover, the present is a linear study, which is rigorously valid only for infinitesimal perturbations.
 If the perturbations are small but finite, we expect our framework to be still in excellent agreement with the real picture.
 However, it is evident that this check can be done only numerically, by performing simulations of the full problem,
 in order to understand whether the nonlinear coupling in the Navier--Stokes advection term induces significant modifications \cite{DCP14c}.
 This issue could e.g.\ be attacked by means of appropriate Large-Eddy Simulations and provide the basis for a more direct comparison with experiments
 \cite{MSC09}.\\
 Last, but definitely not least, a relevant question arises about the stability of the considered steady flows to perturbations
 involving large spatial scales. Further weakly-nonlinear analysis (as in \cite{GVF94} for hydrodynamic flows and in \cite{CZ15} for MHD flows)
 may reveal a complex dynamics of the large-scale perturbations affecting the performance of the vehicle.

 \paragraph{Acknowledgements} We thank Michel Pognant and Dominique Foug\`ere for technical assistance.
 This research was supported by: LabEx ``M\'ecanique et Complexit\'e'' (AMU, France);
 CMUP (UID/MAT/00144/2019), funded by FCT with national (MCTES) and European structural funds through the programs FEDER
 under the partnership agreement PT2020; Project STRIDE - NORTE-01-0145-FEDER-000033, funded by ERDF - NORTE 2020;
 and Project MAGIC - POCI-01-0145-FEDER-032485, funded by FEDER via COMPETE 2020 - POCI and by FCT/MCTES via PIDDAC.

 \appendix
 \section{Details about boundary conditions and adjoint equations} \label{app}

 The boundary conditions for the full velocity field $\bm{u}=\left(\begin{array}{c}u\\v\\w\end{array}\right)$ are the following:
 \begin{itemize}
  \item inlet on segment ED: $u=1$, $v=w=0$;
  \item outlet on segment BC: $p\mathtt{I}-\nu\bm{\nabla}\bm{u}=\mathtt{0}$;
  \item free slip on segments EI and DC: $v=0$, $\nabla_yu=\nabla_yw=0$;
  \item no slip on segments IO, OA and AB: $u=v=w=0$.
 \end{itemize}
 The base flow $\bm{U}$ inherits the same exact conditions. On the contrary, the perturbation field must satisfy fully-homogeneous boundary conditions,
 so that all the formulae above hold also for the quantities with a prime except for the very first one, which becomes (at inlet ED) $u'=0$.

 Equation (\ref{dirper}) can be rewritten in terms of the field $\displaystyle\binom{\bm{u}''}{p''}(x,y)$ as:
 \begin{equation*}
  \left\{\!\!\!\!\begin{array}{rcl}
   \sigma u''+(u''\nabla_x+v''\nabla_y)U+(U\nabla_x+V\nabla_y)u''\!\!&\!\!=\!\!&\!\!-\nabla_xp''+\nu(\nabla_x^2+\nabla_y^2-\beta^2)u''\!\!\\
   \sigma v''+(u''\nabla_x+v''\nabla_y)V+(U\nabla_x+V\nabla_y)v''\!\!&\!\!=\!\!&\!\!-\nabla_yp''+\nu(\nabla_x^2+\nabla_y^2-\beta^2)v''\!\!\\
   \sigma w''+(U\nabla_x+V\nabla_y)w''\!\!&\!\!=\!\!&\!\!-\ui\beta p''+\nu(\nabla_x^2+\nabla_y^2-\beta^2)w''\!\!\\
   \nabla_xu''+\nabla_yv''+\ui\beta w''\!\!&\!\!=\!\!&\!\!0\end{array}\right.
 \end{equation*}
 (with $\bm{u}''=\left(\begin{array}{c}u''\\v''\\w''\end{array}\right)$).

 The procedure to find the adjoint equations consists in performing the scalar product of equations (\ref{dirper}) (considered as a hypervector)
 with the adjoint field $\displaystyle\binom{\bm{\uu}^{\dag}}{\mathcal{P}^{\dag}}$, and in integrating by parts on the whole domain, exploiting the
 convenience of our boundary conditions. In particular, the ones for the adjoint variables are the same as for the direct counterpart, except for the
 outlet condition which reads (on segment BC): $\mathcal{P}^{\dag}\mathtt{I}+\nu\bm{\nabla}\bm{\uu}^{\dag}+\bm{U}\otimes\bm{\uu}^{\dag}=\mathtt{0}$.

 However, when a forcing term is also present in the direct equations (as in (\ref{forper})), the procedure is more complex because it involves the whole
 formalism of Lagrange multipliers and functional derivatives \cite{P09,M07}. We do not report it here, and we simply remind the final result
 (\ref{adjper}).

\end{document}